\documentclass[twocolumn,prb,aps,superscriptaddress]{revtex4}
\usepackage{graphicx}
\usepackage{color}
\usepackage{amsmath}
\usepackage{cases}
\newcommand{\be}{\begin{equation}}
\newcommand{\ee}{\end{equation}}
\newcommand{\bea}{\begin{eqnarray}}
\newcommand{\eea}{\end{eqnarray}}
\newcommand{\bse}{\begin{subequations}}
\newcommand{\ese}{\end{subequations}}

\begin{document}
\title{Crystallographic, electronic, thermal and magnetic properties of single-crystal SrCo$_2$As$_2$}
\author{Abhishek Pandey}
\altaffiliation{apandey@ameslab.gov} 
\author{D. G. Quirinale}
\author{W. Jayasekara}
\author{A. Sapkota}
\affiliation {Ames Laboratory and Department of Physics and Astronomy, Iowa State University, Ames, Iowa 50011, USA}
\author{M. G. Kim}
\affiliation {Ames Laboratory and Department of Physics and Astronomy, Iowa State University, Ames, Iowa 50011, USA}
\affiliation{Materials Sciences Division, Lawrence Berkeley National Laboratory, Berkeley, California 94720, USA}
\author{R. S. Dhaka}
\altaffiliation {Present address: Swiss Light Source, Paul Scherrer Institute, CH-5232 Villigen PSI, Switzerland}
\author{Y. Lee}
\affiliation{Ames Laboratory and Department of Physics and Astronomy, Iowa State University, Ames, Iowa 50011, USA}
\author{T. W. Heitmann}
\affiliation{The Missouri Research Reactor, University of Missouri, Columbia, Missouri 65211, USA}
\author{P. W. Stephens}
\affiliation{Department of Physics and Astronomy, SUNY at Stony Brook, Stony Brook, New York 11794, USA}
\author{V. Ogloblichev}
\affiliation{Institute of Metal Physics, Ural Division of Russian Academy of Sciences, Ekaterinburg 620990, Russia}
\author{A.~Kreyssig}
\author{R. J. McQueeney}
\author{A. I. Goldman}
\author{Adam Kaminski}
\author{B. N. Harmon}
\author{Y. Furukawa}
\author{D. C. Johnston} 
\altaffiliation{johnston@ameslab.gov}
\affiliation{Ames Laboratory and Department of Physics and Astronomy, Iowa State University, Ames, Iowa 50011, USA}

\date{July 24, 2013}

\begin{abstract}

In tetragonal $\rm{SrCo_2As_2}$ single crystals, inelastic neutron scattering measurements demonstrated that strong stripe-type antiferromagnetic (AFM) correlations occur at a temperature $T = 5$~K [W. Jayasekara {\it et al.}, arXiv:1306.5174] that are the same as in the isostructural $A{\rm Fe_2As_2}$ ($A$ = Ca, Sr, Ba) parent compounds of high-$T_{\rm c}$ superconductors.  This surprising  discovery suggests that $\rm{SrCo_2As_2}$ may also be a good parent compound for high-$T_{\rm c}$ superconductivity.  Here, structural and thermal expansion, electrical resistivity $\rho$, angle-resolved photoemission spectroscopy (ARPES), heat capacity $C_{\rm p}$, magnetic susceptibility $\chi$, $^{75}$As NMR and neutron diffraction measurements of $\rm{SrCo_2As_2}$ crystals are reported together with LDA band structure calculations that shed further light on this fascinating material.  The $c$-axis thermal expansion coefficient $\alpha_c$ is {\it negative} from 7 to 300~K, whereas $\alpha_a$ (the $a$-axis thermal expansion coefficient) is positive over this $T$ range. The $\rho(T)$ shows metallic character.  The ARPES measurements and band theory confirm the metallic character and in addition show the presence of a flat band near the Fermi energy $E_{\rm F}$.  The band calculations exhibit an extremely sharp peak in the density of states ${\cal D}(E\approx E_{\rm F})$ arising from a flat $d_{x^2-y^2}$ band, where the $x$ and $y$ axes are along the $a$ and $b$ axes of the Co square lattice, respectively.  A comparison of the Sommerfeld coefficient of the electronic specific heat with $\chi(T\to0)$ suggests the presence of strong ferromagnetic itinerant spin correlations which on the basis of the Stoner criterion predicts that $\rm{SrCo_2As_2}$ should be an itinerant ferromagnet, in conflict with the magnetization data.  The $\chi(T)$ does have a large magnitude, but also exhibits a broad maximum at $\approx 115$~K suggestive of dynamic short-range AFM spin correlations, in agreement with the neutron scattering data.  The measurements show no evidence for any type of phase transition between~1.3~and 300~K and we suggest that metallic $\rm{SrCo_2As_2}$ has a gapless quantum spin-liquid ground state.

\end{abstract}


\maketitle

\section{Introduction}

After the discovery of high-temperature superconductivity (SC) below $T_{\rm c} = 26$~K in tetragonal LaFeAsO$_{1-x}$F$_x$ (1111-type) in 2008,\cite{Kamihara2008} a worldwide effort began to understand the mechanism for the SC and to discover other Fe-based superconductors.  This effort quickly led to the enhancement of $T_{\rm c}$ to its current record among bulk Fe-based superconductors of $T_{\rm c} = 56.3$~K for the 1111-type compound ${\rm Gd_{0.8}Th_{0.2}FeAsO}$.\cite{Wang2008}  Additional families of Fe-based high-$T_{\rm c}$ superconductors were soon discovered including the 122-type arsenides where the parent compounds have composition $A{\rm Fe_2As_2}$ ($A$ = Ca, Sr, Ba, Eu)  with the tetragonal ${\rm ThCr_2Si_2}$-type structure (space group $I4/mmm$).\cite{Johnston2010, Paglione2010, Stewart2011}  All families of Fe-based high-$T_{\rm c}$ superconductors contain stacked square lattices of Fe atoms that are identical to the stacked square lattices of Cu atoms in the layered high-$T_{\rm c}$ cuprates.\cite{Johnston2010,Bednorz1986, Johnston1997, Kastner1998}

The 122-type family of iron arsenide-based superconductors is the most studied because large, high-quality single crystals can be grown,\cite{Canfield2010} and the parent compounds offer enormous flexibility for chemical substitutions on the $A$, Fe and As sites.  Crystals with masses $\sim 1$~g can be grown, which allows a broad range of measurements including inelastic neutron scattering measurements to be performed to characterize and understand the properties.  It has been demonstrated that the parent compounds at ambient pressure exhibit a lattice distortion to an orthorhombic structure at a system-dependent $T_{\rm S}\lesssim 200$~K together with long-range antiferromagnetic (AFM) ordering at the same or somewhat lower N\'eel temperature $T_{\rm N}$.\cite{Johnston2010, Lumsden2010}  The ordering propagation vector is the same as the nesting wave vector between the hole and electron pockets of the semimetallic band structure.  This and the low ordered moments\cite{Johnston2010, Lumsden2010} $\mu\lesssim 1\,\mu_{\rm B}$/Fe indicate an itinerant origin of the AFM\cite{Mazin2008,Dong2008} rather than arising from interactions between local Fe magnetic moments.

The AFM structure within the $ab$-plane of the Fe square lattice is a so-called ``stripe'' structure in which ferromagnetically aligned stripes of Fe magnetic moments along the orthorhombic $b$ axis are AFM aligned along the $a$~axis.\cite{Johnston2010, Lumsden2010} These orthorhombic axes have the same directions as the respective Fe square lattice $a$ and $b$ axes which are in turn rotated by 45$^\circ$ with respect to the $a$ and~$b$ axes of the body-centered tetragonal ${\rm ThCr_2Si_2}$-type crystal structure.  Furthermore, in order for high-$T_{\rm c}$ SC to occur, this long-range AFM ordering must be largely suppressed by chemical substitutions or pressure, but strong dynamic short-range AFM correlations must still be preserved.  These AFM fluctuations/correlations are widely considered to be the glue for the SC Cooper pairs.\cite{Johnston2010, Mazin2008, Mazin2009, Fernandes2010}  A similar phenomenology and  temperature-doping phase diagram are found for the high-$T_{\rm c}$ cuprates.\cite{Bednorz1986, Johnston1997, Kastner1998, Basov2011}

Among the substituted 122-type materials,\cite{Johnston2010} the best studied are Ba$_{1-x}$K$_x{\rm Fe_2As_2}$ with $T_{\rm c}\leq 38$~K,\cite{Rotter2008}  Ba(Fe$_{1-x}$Co$_x)_2{\rm As_2}$ with $T_{\rm c}\leq 25$~K,\cite{Sefat2008, Leithe-Jasper2008, Nandi2010} and BaFe$_2$(As$_{1-x}$P$_x)_2$ with $T_{\rm c}\leq 30$~K.\cite{Jiang2009}   The detailed temperature-composition phase diagram for Ba(Fe$_{1-x}$Co$_x)_2{\rm As_2}$ with $0\leq x \leq 0.12$ shows, as $T\to0$, an AFM region at $0\leq x \lesssim 0.03$, a region of coexisting AFM and SC for $0.03 \lesssim x \lesssim 0.06$ and a region of only SC order (the remaining part of the superconducting dome) for $0.06\lesssim x \lesssim 0.12$, where $T_{\rm c}\to0$ at $x\approx0.12$.\cite{Nandi2010}  Inelastic neutron scattering measurements demonstrated that the AFM correlations/fluctuations that are present in the paramagnetic (PM) state of ${\rm BaFe_2As_2}$ above $T_{\rm N} = 136$~K (Ref.~\onlinecite{Matan2009}) are absent in the nonsuperconducting PM composition $x = 0.14$ at 10~K,\cite{Sato2011} consistent with the hypothesis that AFM spin fluctuations are required as the SC pairing glue in these materials.\cite{Johnston2010, Lumsden2010}

\begin{figure}[t]
\includegraphics[width=2.3in,viewport= 10 0 330 445,clip]{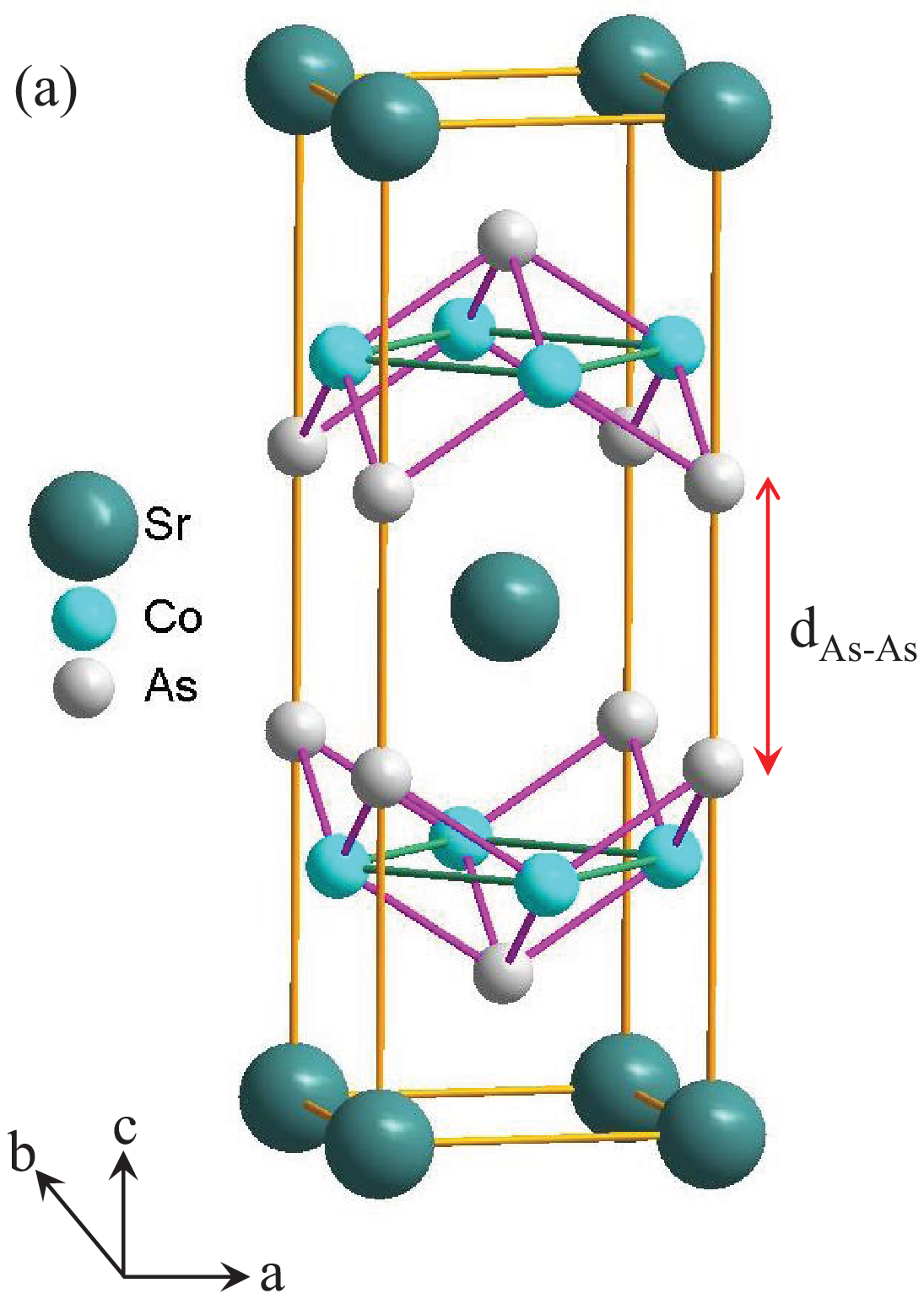}\vspace{0.2in}
\includegraphics[width=3.in]{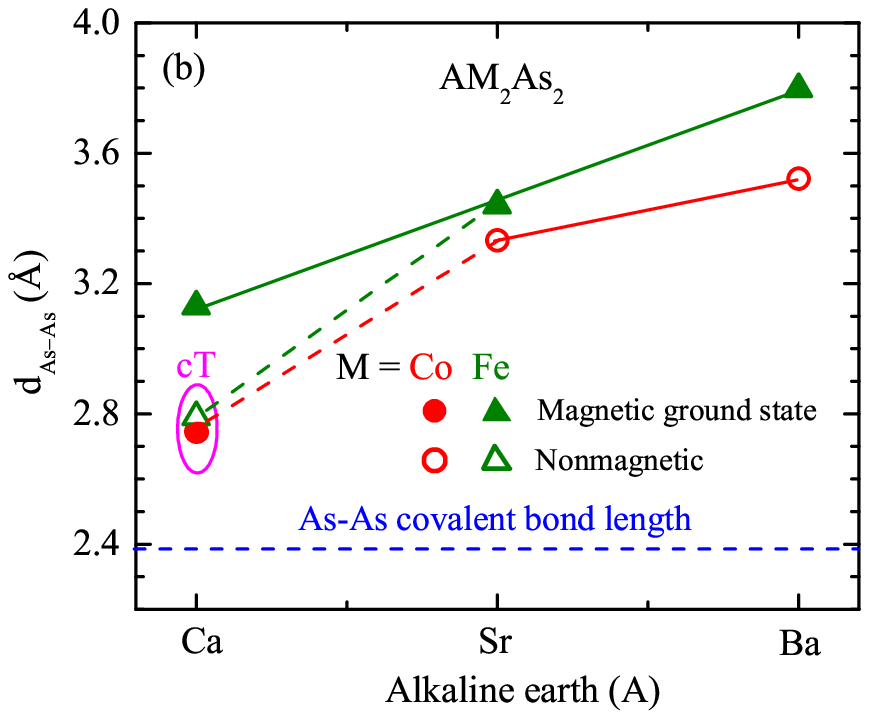}
\caption{(Color online) (a) ${\rm ThCr_2Si_2}$-type body-centered tetragonal crystal structure of ${\rm SrCo_2As_2}$ with space group $I4/mmm$. The interlayer As--As distance $d_{\rm As-As}$ is indicated.  (b) $d_{\rm As-As}$ versus $A$ atom in $A\rm{Co_2As_2}$ and $A\rm{Fe_2As_2}$ ($A$ = Ca, Sr, Ba).\cite{Anand2012b}  The lines are guides to the eye. The horizontal dashed line indicates the As--As single covalent bond distance of 2.38~\AA.\cite{Cordero2008}  }
\label{fig:Figure_Lattice}
\end{figure}

As might be expected from the latter result, neither magnetic susceptibility\cite{Leithe-Jasper2008} $\chi(T)$ nor neutron diffraction\cite{Jayasekara2013} measurements on ${\rm SrCo_2As_2}$ showed any evidence for long-range magnetic ordering above 1.8 and 4~K, respectively.  This is the same result as found for isostructural ${\rm BaCo_2As_2}$,\cite{Sefat2009} but is in contrast to the magnetic behavior of ${\rm CaCo_2As_2}$ which shows A-type AFM ordering below a sample-dependent $T_{\rm N}$ of~53 to 76~K.\cite{Cheng2012, Ying2012, Anand2013}  The ${\rm ThCr_2Si_2}$-type crystal structure of ${\rm SrCo_2As_2}$ (Ref.~\onlinecite{Pfisterer1980}) is shown in Fig.~\ref{fig:Figure_Lattice}(a).

The As--As interlayer distance $d_{\rm As-As}$ [see Fig.~\ref{fig:Figure_Lattice}(a)] is correlated with the magnetic ground state,\cite{Anand2012b} where a ``layer'' is defined as an As--$M$--As layer and $M$ is the transition metal, as shown in Fig.~\ref{fig:Figure_Lattice}(b).  ${\rm CaCo_2As_2}$ with an AFM ground state has a collapsed tetragonal (cT) structure reflecting approach to interlayer As--As dimerization, where the As--As single-covalent-bond distance\cite{Cordero2008} is indicated in Fig.~\ref{fig:Figure_Lattice}(b), whereas ${\rm BaCo_2As_2}$ with a PM ground state has an uncollapsed tetragonal (T) structure without such dimerization.  ${\rm SrCo_2As_2}$ also has a PM ground state and is in an intermediate regime,\cite{Anand2012b} as seen in Fig.~\ref{fig:Figure_Lattice}(b).  This correlation between the magnetic behavior and the crystal structure contrasts with that of the 122-type $ A{\rm Fe_2As_2}$ compounds where the cT structure of ${\rm CaFe_2As_2}$ has a nonmagnetic ground state and the T structures of the Ca, Sr and Ba compounds all have AFM ground states.\cite{Johnston2010,Anand2012b, Kreyssig2008, Goldman2009, Pratt2009}  Interlayer As--As dimerization in the cT phase results in a valence of As$^{-2}$, i.e.\ [As--As]$^{-4}$,  instead of the usual As$^{-3}$ valence in the 122-type (Ca,\,Sr,\,Ba)$M_2$As$_2$ compounds, corresponding to a formal $M$ oxidation state of $M^{+1}$ instead of the usual $M^{+2}$.  From Fig.~\ref{fig:Figure_Lattice}(b), ${\rm SrCo_2As_2}$ may be in a regime with intermediate Co oxidation state between Co$^{+1}$ in ${\rm CaCo_2As_2}$ and Co$^{+2}$ in ${\rm BaCo_2As_2}$.  This intermediate valence state, in turn, might be expected to result in strong dynamic Co spin fluctuations in ${\rm SrCo_2As_2}$. 

In view of the disappearance of dynamic AFM correlations in the PM state of the Ba(Fe$_{1-x}$Co$_x)_2{\rm As_2}$ system as $x$ increases from $x=0$ to $x=0.14$,\cite{Matan2009, Sato2011} it is amazing that strong AFM stripe-type correlations are present in the $x=1$ end-member ${\rm SrCo_2As_2}$ of the similar series Sr(Fe$_{1-x}$Co$_x)_2{\rm As_2}$ (Refs.~\onlinecite{Hu2011, Zhao2008, Ewings2011}) as established from inelastic neutron scattering measurements on single crystals.\cite{Jayasekara2013}  The above-noted intermediate valence of the Co in ${\rm SrCo_2As_2}$ may be responsible for the lack of long-range AFM order, at least in part.  The presence of these AFM correlations in ${\rm SrCo_2As_2}$ raises key questions about how they come about in this compound, why this compound is not superconducting, and whether it could be made superconducting by some type of chemical substitution or by application of pressure.

Here we report the investigation of undoped ${\rm SrCo_2As_2}$ single crystals using x-ray and neutron diffraction, in-plane ($ab$-plane) electrical resistivity~$\rho$, angle-resolved photoemission spectroscopy (ARPES), heat capacity $C_{\rm p}$, magnetization $M$, $\chi$ and $^{75}$As nuclear magnetic resonance (NMR) measurements, complemented by LDA band structure calculations.  Experimental and theoretical details are given in Sec.~\ref{Sec:ExpDetails}.  The x-ray diffraction investigations of crushed crystals in Sec.~\ref{Sec:Crystallography} yield a structure in agreement with previous work.  The $\rho(T)$ measurements in Sec.~\ref{Sec:rho} demonstrate metallic character, consistent with the ARPES measurements in Sec.~\ref{Sec:ARPES} which also indicate that there is no obvious significant Fermi surface nesting.  The band theory calculations in Sec.~\ref{Sec:BandTheory} reveal a sharp and high peak in the density of states near the Fermi energy ${\cal D}(E\approx E_{\rm F})$.  The $C_{\rm p}$ measurements in Sec.~\ref{Sec:Cp} further confirm the metallic character and confirm the presence of a large ${\cal D}(E\approx E_{\rm F})$.  The $M$ and $\chi$ measurements in Sec.~\ref{Sec:M_chi} show clear evidence for dynamic short-range AFM spin correlations in a quasi-low-dimensional spin lattice, consistent with the inelastic neutron scattering results.\cite{Jayasekara2013}  The NMR measurements are presented in Sec.~\ref{Sec:NMR} which indicate the presence of stripe-type and/or FM correlations.  The x-ray diffraction, $\rho$, $C_{\rm p}$, $\chi$ and NMR measurements reveal no obvious phase transitions between 1.3 and 300~K, in agreement with the previous\cite{Jayasekara2013} and present neutron diffraction measurements in Sec.~\ref{Sec:MagneticNeuts}.  In Sec.~\ref{Sec:GndState} we discuss the ground state of metallic ${\rm SrCo_2As_2}$ and suggest that this compound has a gapless quantum spin-liquid ground state.  A summary and our conclusions are given in Sec.~\ref{Sec:Summary}. 

\section{\label{Sec:ExpDetails} Experimental and Theoretical Details}

Single crystals of $\rm{SrCo_2As_2}$ were grown from solution using Sn flux.  In the Fe-based high-$T_{\rm c}$ superconductivity field, Sn flux was first used by Ni {\it et al.}\ to grow crystals of ${\rm BaFe_2As_2}$ and ${\rm Ba_{0.55}K_{0.45}Fe_2As_2}$.\cite{Ni2008}  High-purity elements Sr (99.95\%) from Sigma-Aldrich, and Co (99.998\%), As (99.99999\%) and Sn (99.999\%) from Alfa Aesar, were taken in 1:2:2:45 ratio and placed in an alumina crucible that was sealed in an evacuated silica tube. After prereacting the elements at 610~$^{\circ}$C for 5~h, the assembly was heated to 1150~$^{\circ}$C, left there for 8~h and then slowly cooled to 700~$^{\circ}$C in 160~h. At this temperature the flux was decanted using a centrifuge. The typical size of the crystals obtained was $2 \times 2 \times 0.3$~mm$^3$. However, by increasing the size of the crucible to a 10~mL capacity, large crystals of $\rm{SrCo_2As_2}$ with masses up to $\approx 1$~g with dimensions up to $15\times10\times1\ {\rm mm^3}$ can be grown out of Sn flux.\cite{Jayasekara2013}

The chemical compositions of the crystals were determined by energy-dispersive x-ray analysis (EDX) using a JEOL-JSM-5910LV scanning electron microscope and found to be in agreement with the expected stoichiometric composition $\rm{SrCo_2As_2}$. These EDX measurements showed no visible peaks at the expected positions of the Sn x-ray emission lines, and the amount of Sn incorporated into the crystals from the Sn flux was given by the software as 0.22(16)~at\%, which is consistent with zero to within $1.3\sigma$.  However, we cannot rule out the possibility that a very small amount of Sn was incorporated into the crystals which might have had some influence on the results of our physical property measurements.  For comparison, $\rm{BaFe_2As_2}$ crystals grown in Sn flux contain $\sim1$~at\% Sn as determined using wavelength-dispersive x-ray analysis,\cite{Ni2008} which results in large changes in the physical properties such as in a reduction of the tetragonal to orthorhombic structural transition temperature from 136~K in pure crystals grown out of FeAs self-flux as discovered by Wang~{\it et al.}\cite{Wang2009} to the value of 85~K in the crystals grown out of Sn flux by Ni~et~al.\cite{Ni2008}

Structural characterization was performed using room-temperature powder x-ray diffraction (XRD) data obtained on crushed crystals using a Rigaku Geigerflex powder diffractometer and Cu-$K_{\alpha}$ radiation.  Rietveld refinement of the XRD data was carried out using the {\tt FULLPROF} package.\cite{Carvajal1993}  The $M(H)$, $\chi$, $C_{\rm p}$ and four-probe $\rho$ measurements were performed using instruments from Quantum Design, Inc.  In the $M(H)$, $\chi$ and $C_{\rm p}$ measurements, the contribution of the sample holder was measured separately and corrected for in each case.

A high-resolution x-ray powder diffraction pattern was collected at ambient temperature at the beamline X16C at the National Synchrotron Light Source.  The sample was ground with an agate mortar and pestle in a glovebox having a water vapor concentration of a few parts per million, loaded into a glass Lindemann capillary of 0.3-mm nominal diameter, and flame-sealed without exposure to the atmosphere. The x-ray wavelength, 0.6995~\AA, was chosen using a Si(111) double monochromator. The powder diffraction pattern was collected in the 5--45$^\circ$ 2$\theta$ range with a constant step size of 0.005$^\circ$, and a linearly varying counting time of \mbox{1--3~s/point}. The incident beam intensity was monitored with an ion chamber and the diffracted radiation was measured with a NaI(Tl) scintillation detector. The axial and in-plane resolution of the diffractometer were set by slits and a Ge(111) analyzer crystal, respectively.  Rietveld refinement of the x-ray diffraction pattern was done using the GSAS\cite{Larson2004} package.

High-resolution single-crystal x-ray diffraction measurements were performed on a four-circle diffractometer using Cu-$K_{\alpha1}$ radiation from a rotating anode x-ray source, selected by a Ge(111) monochromator.  For measurements of the temperature dependence of the lattice parameters, a platelike single crystal, with the tetragonal \emph{c}-axis perpendicular to the plate, was attached to a flat copper sample holder on the cold finger of a He closed-cycle displex refrigerator.  The sample was oriented such that the ($h$~0~$\ell$) plane or the $(h~h~\ell)$ plane of the reciprocal lattice was coincident with the scattering plane of the diffractometer.

Single-crystal neutron diffraction measurements were done on the TRIAX triple-axis spectrometer at the University of Missouri Research Reactor employing an incident neutron energy of 14.7~meV\@. A plate-like crystal of mass 302.4~mg was mounted on the cold finger of a closed-cycle displex refrigerator with the ($h~0~\ell$) plane of the reciprocal lattice coincident with the scattering plane of the diffractometer, allowing us to search for magnetic scattering at the reciprocal lattice positions for the A-type AFM structure [($h~0~\ell$) with $h + \ell = 2n+1$], as found previously for CaCo$_2$As$_2$,\cite{Anand2013} and the G-type AFM structure [($h~0~\ell$) with~$h$ and~$\ell$ odd], as found for the related compound BaMn$_2$As$_2$.\cite{SinghY2009}

High-resolution ARPES experiments were carried out using a Scienta R4000 electron analyzer and 140~eV photon energy at beamline~7.0.1 of the Advanced Light Source (ALS), Berkeley, California.  The crystals were cleaved {\it in situ} along the {\it ab} plane and measured at a pressure of $\approx4\times10^{-11}$~mbar.  The energy and momentum resolution were set to $\sim20$~meV and $\sim0.3^{\circ}$, respectively. The Fermi energy ($E_{\rm F}$) of the sample was referenced to that of a gold sample deposited {\it in situ} on the sample holder.  In our notation the Brillouin zone (BZ) of ${\rm SrCo_2As_2}$ is that of the body-centered-tetragonal ($I4/mmm$) direct lattice  and not that of the Co square lattice, and the $(k_x,k_y,k_z)$ axes of the Brillouin zone are, as conventionally defined in the literature, in the directions of the direct lattice translation vectors ({\bf a},{\bf b},{\bf c}), respectively.\cite{Johnston2010}

Nuclear magnetic resonance (NMR) measurements were carried out on $^{75}$As (\textit{I} = 3/2, $\gamma_{\rm N}/2\pi$ = 7.2919~MHz/T)  using a homemade phase-coherent spin-echo pulse spectrometer.  $^{75}$As-NMR spectra were obtained by sweeping the magnetic field $H$ at a fixed frequency $f = 51.1$~MHz.  The magnetic field was applied parallel to either the crystal $c$ axis or the $ab$-plane.  The origin of the NMR shift $K = 0$ of the $^{75}$As nucleus was determined by $^{75}$As NMR measurements of GaAs.  The $^{75}$As nuclear spin-lattice relaxation rate (1/$T_{\rm 1}$) was measured with a saturation recovery method. 

For electronic structure calculations, we used the full-potential linear augmented plane wave (FP-LAPW)\cite{Blaha2001} method with the local density approximation (LDA).\cite{Perdew1992} To obtain self consistent charge and wave functions,  we employed ${R_{\rm MT}}\times k_{\rm max}=9.0$  with muffin tin radii ${R_{\rm MT}} =  2.3,$ 2.1, and 2.1~a.u.\ for Sr, Co and As, respectively. 828 {\bf k}~points were selected in the irreducible Brillouin zone and calculations were iterated to reach the total energy convergence criterion which was 0.01~mRy/cell, where a ``cell'' is a primitive cell containing one formula unit of five atoms.  Our experimental lattice parameters in Table~\ref{Table:Structure} for the $I4/mmm$ structure of ${\rm SrCo_2As_2}$ ($a=3.9471$~\AA, $c=11.801$~\AA) were used in the calculations.  To obtain the theoretical $c$-axis position $z_{\rm As}$ of the As atom, its position was relaxed until the force on the As atom was smaller than 1~mRy/a.u., which gave $z_{\rm As}=0.35146$. This value is somewhat smaller than our experimental value $z_{\rm As} = 0.3588(1)$ in Table~\ref{Table:Structure}.

\section{\label{Sec:Crystallography} Crystallography and Thermal Expansion}

\subsection{Powder X-Ray Diffraction: Crystal Structure}

\begin{figure}[t]
\includegraphics[width=3.3in]{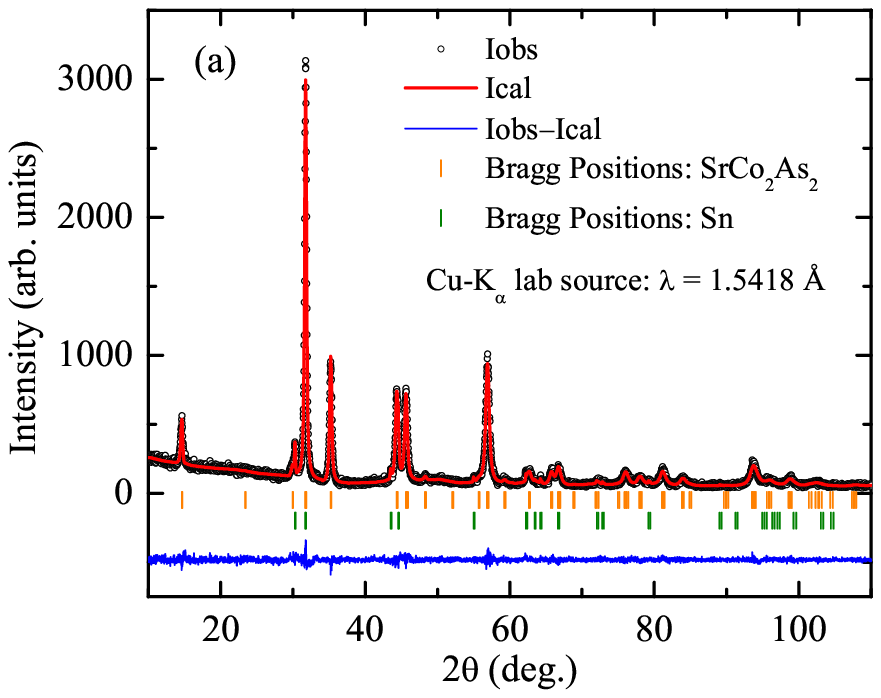}\vspace{0.1in}
\includegraphics[width=3.3in]{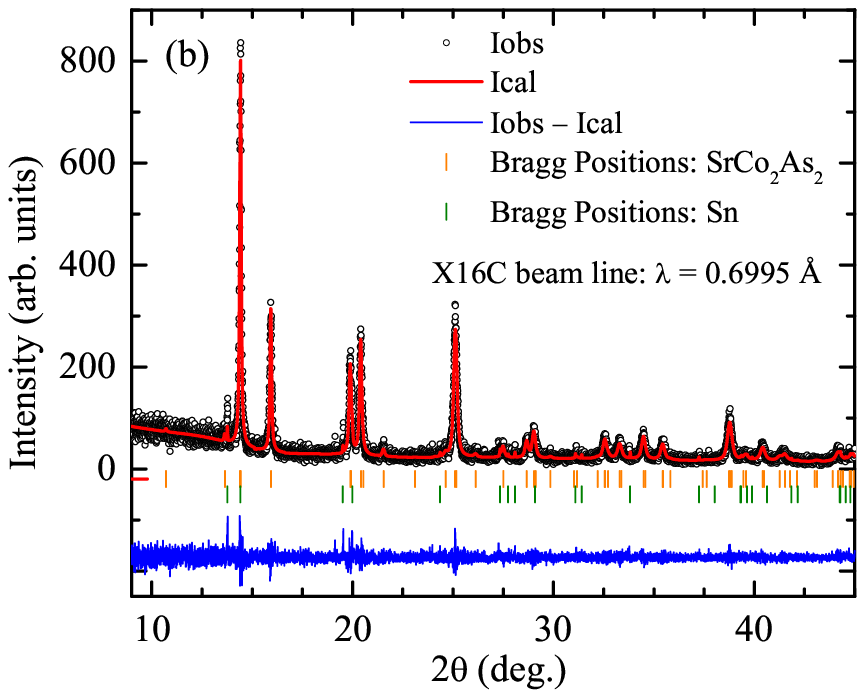}
\caption{(Color online) Room-temperature powder XRD data for $\rm{SrCo_2As_2}$ along with the Rietveld refinement fit, Bragg peak positions and difference profile from (a) a laboratory-based diffractometer with x-ray wavelength $\lambda = 1.5418$~\AA\ and (b) a synchrotron-based diffractometer with x-ray wavelength $\lambda = 0.6995$~\AA. A two-phase Rietveld refinement was performed to account for the peaks from a minor amount of Sn flux on the surfaces of each of the two crushed-crystal samples.}
\label{fig:Figure_XRD}
\end{figure}

\begin{table*}
\caption{Crystal data for $A\rm{Co_2As_2}$ ($A =$ Ca, Sr, Ba) compounds that crystallize in the $\rm {ThCr_2Si_2}$--structure with $I4/mmm$ space group symmetry and $Z=2$ formula units/unit cell.  The $A$ atoms are at the $2a$~(0~0~0) sites, the Co atoms at the $4d$~(0~$\frac{1}{2}$~$\frac{1}{4}$) sites and the As atoms at the $4e$~(0~0~$z_{\rm As}$) sites.  For the laboratory-based measurements, the goodness of fit parameters are $\chi^2 = 1.19$, $R_{\rm p} = 7.08$\% and $R_{\rm wp} = 9.34$\%, whereas for the synchrotron-based measurements they are $\chi^2 = 1.08$, $R_{\rm p} = 12.47$\% and $R_{\rm wp} = 16.67$\%. The designations ``T'' and ``cT'' refer to the uncollapsed tetragonal and collapsed tetragonal structures, respectively.  ${\rm SrCo_2As_2}$ has $c/a$ and $d_{\rm As-As}$ values intermediate between those of T-type~${\rm BaCo_2As_2}$ and cT-type~${\rm CaCo_2As_2}$.}
\label{Table:Structure}
\begin{ruledtabular}
\begin{tabular}{l c c l l c c c}
Compound & $a$ (\AA)	& $c$ (\AA)	& $c/a$ & $z_{\rm As}$ & $d_{\rm As-As}$ (\AA) & Structure & Reference\\

\hline
$\rm {CaCo_2As_2}$ & 3.9831(2) & 10.2732(6) & 2.5792(3) & 0.3664(2) & 2.745(4) & cT & [\onlinecite{Anand2012b, Anand2013}] \\
				& 3.989(3) & 10.33(1) & 2.590(9) &	0.372	& 2.64	&&	[\onlinecite{Pfisterer1980}]	\\
$\rm {SrCo_2As_2}$ & 3.9471(4) & 11.801(1)  & 2.9898(6) & 0.3588(1) & 3.333(3) &  & This work\footnotemark[1]\\
				& 3.9466(2) & 11.773(1)  & 2.9831(4) &  0.3587(3) & 3.327(7) &  & This work\footnotemark[2]\\
				& 3.935(7)  & 11.83(2)   & 3.006(10) & 0.362  &  3.27   &   & [\onlinecite{Pfisterer1980}]\\
			$\rm {BaCo_2As_2}$ & 3.958(5)  & 12.67(2)   & 3.201(9) & 0.361  & 3.52    & T  & [\onlinecite{Pfisterer1980}]\\
	\end{tabular}
	\end{ruledtabular}
\footnotetext[1]{Laboratory-based x-ray powder diffraction measurements.}
\footnotetext[2]{Synchrotron-based x-ray powder diffraction measurements.}
\end{table*}

The laboratory-based powder XRD data for ${\rm  SrCo_2As_2}$ are shown in Fig.~\ref{fig:Figure_XRD}(a), along with the two-phase Rietveld refinement fit including a fit to the adventitious Sn from the flux on the surface of the crystals prior to crushing them, the difference profile and Bragg positions for the main phase as well as for Sn.  The parameters fitted during the refinement were the lattice parameters, As $c$-axis position parameter $z_{\rm As}$, isotropic thermal parameters, scale factor, zero shift, three resolution parameters, profile shape parameter and  asymmetry parameters.  The Sn impurity phase fraction was found to be 4.9~wt\%.  The site occupancies of Sr, Co and As were kept fixed at 100\% during the final refinement cycle because allowing the occupancies to vary from this value did not improve the fit significantly.  The lattice parameters and $z_{\rm As}$ are listed in Table~\ref{Table:Structure} along with the corresponding literature data for ${\rm  CaCo_2As_2}$, ${\rm SrCo_2As_2}$ and ${\rm BaCo_2As_2}$.\cite{Anand2012b, Anand2013, Pfisterer1980}  Our data for ${\rm SrCo_2As_2}$ are in reasonable agreement with those of Ref.~\onlinecite{Pfisterer1980}.

The synchrotron x-ray diffraction pattern for ${\rm  SrCo_2As_2}$, the Rietveld profile refinement fit and residuals are displayed in Fig.~\ref{fig:Figure_XRD}(b).  Here again, an impurity phase of adventitious Sn flux, comprising in this sample approximately 1\% of the sample by weight, was included in the refinement of the powder diffraction pattern.  A small section of low angle data was removed from the refinement due to poor statistics.  Several very low intensity peaks could not be identified conclusively with other impurity phases. Table~\ref{Table:Structure} gives a summary of the results of the refinement of the powder XRD data for the $\rm{SrCo_2As_2}$ phase.  These data are in good agreement with those found above from the laboratory-based x-ray diffraction measurements.

\subsection{\label{Sec:ThermalExpansion} Single-Crystal X-ray Diffraction: Thermal Expansion and Search for a Temperature-Induced Orthorhombic Distortion}

\begin{figure}
\includegraphics[width=3.3in]{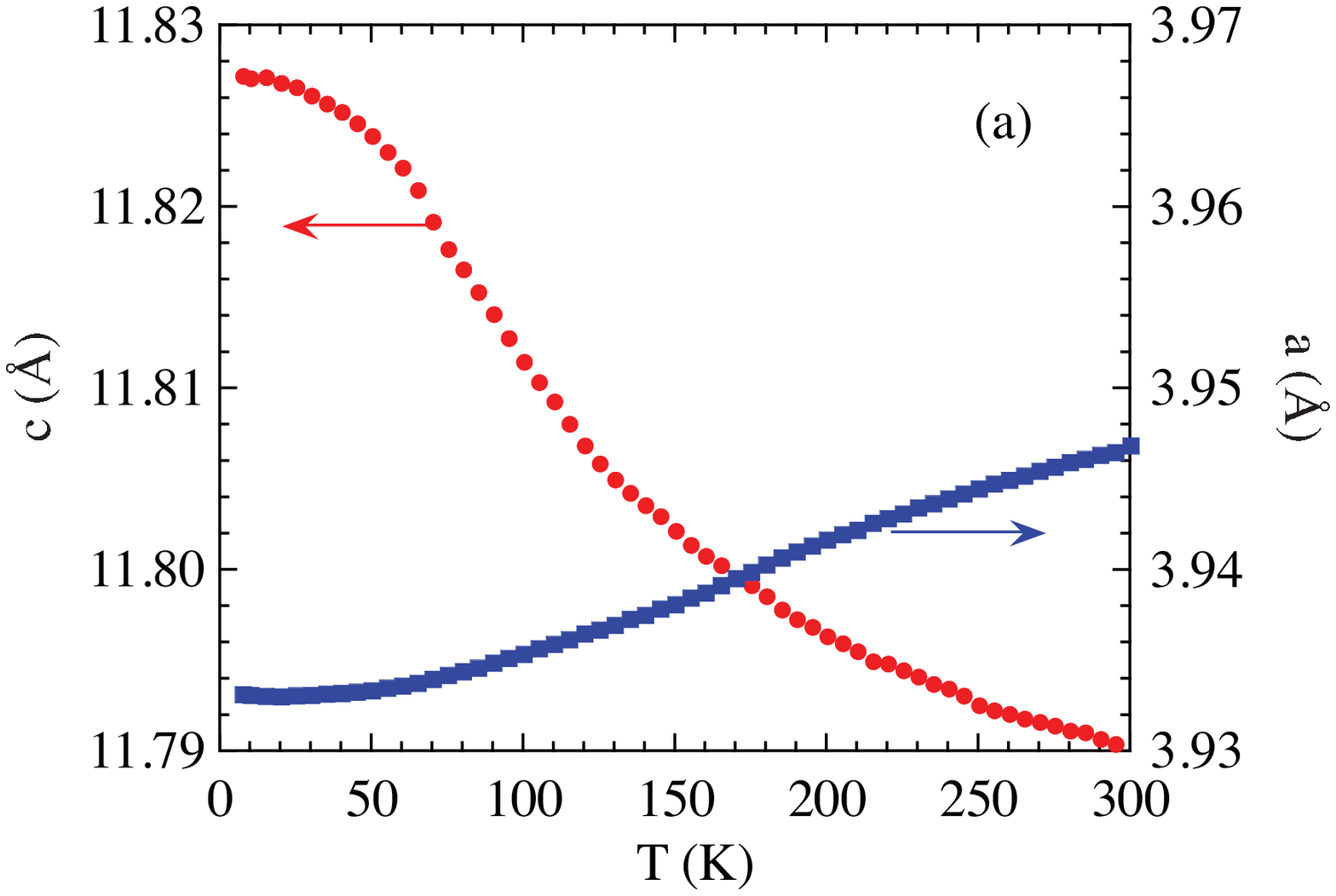}\vspace{-0.in}\vspace{-0.1in}
\includegraphics[width=3.33in,viewport=10 00 520 350,clip]{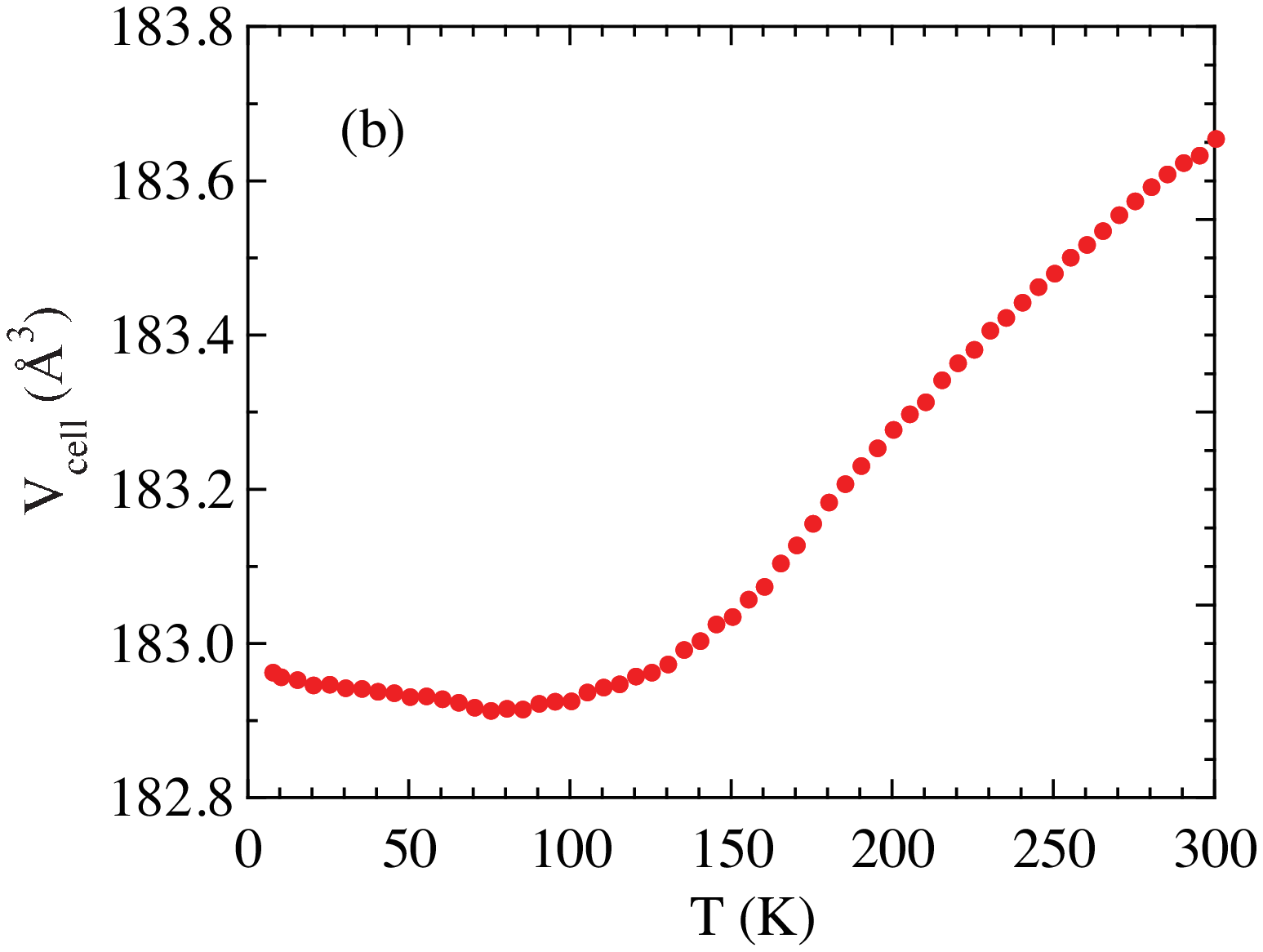}\vspace{-0.1in}
\includegraphics[width=3.45in,viewport=-9 00 520 350,clip]{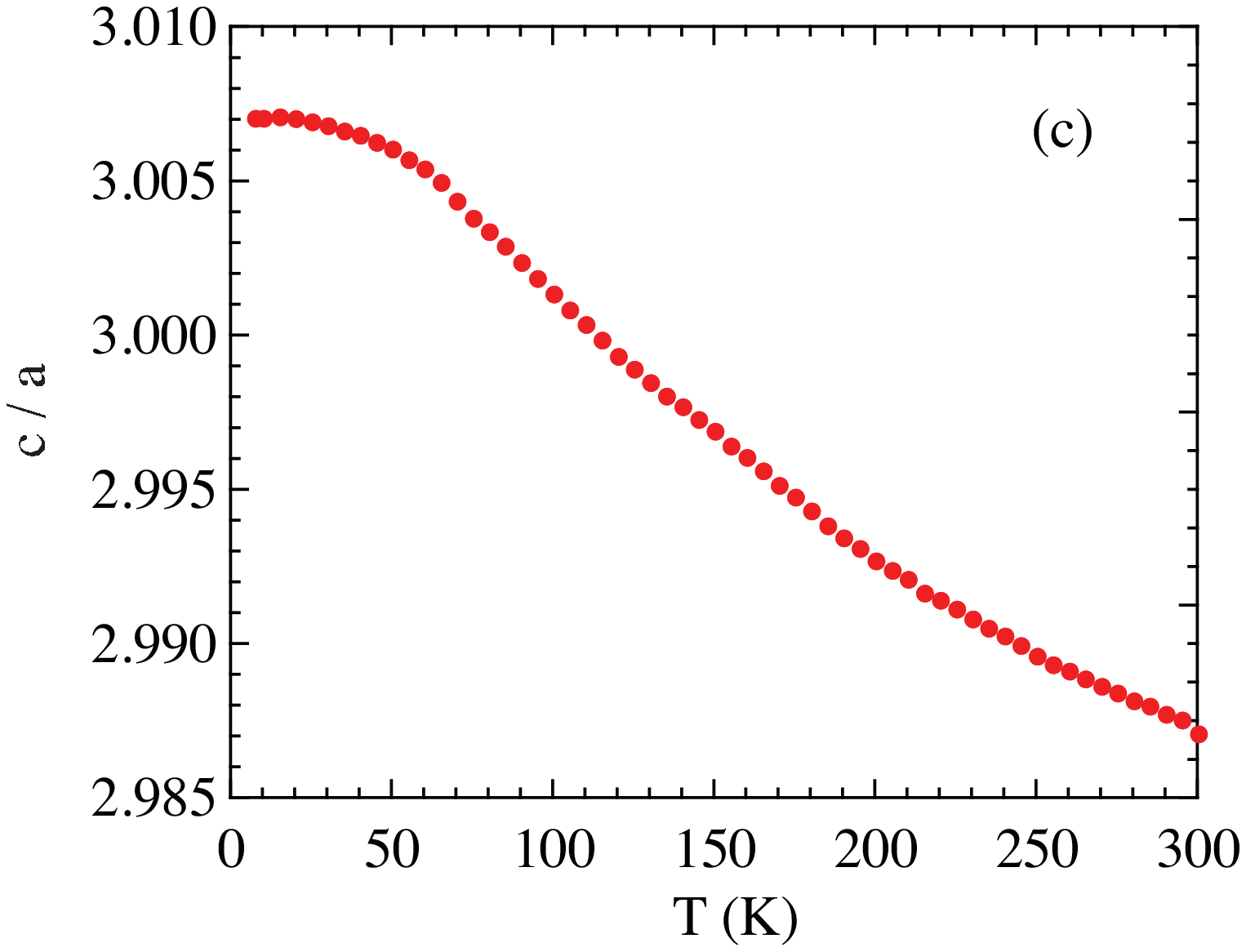}
\caption{(Color online) Temperature dependence of (a) the $a$ and $c$ lattice parameters (right- and left-hand ordinates, respectively), (b) unit cell volume $V_{\rm cell}=a^2c$ and (c) $c/a$ ratio of ${\rm SrCo_2As_2}$ from single-crystal x-ray diffraction measurements.}
\label{Tdep}
\end{figure}

The $T$ dependences of the $a$ and $c$ lattice parameters were extracted from the positions of the (2~0~8) and (0~0~8) Bragg peaks measured by single-crystal \mbox{x-ray} diffraction as described in Sec.~\ref{Sec:ExpDetails}. Figure~\ref{Tdep}(a) shows a striking \emph{decrease} in the $c$~lattice parameter by approximately 0.3\% as $T$ {\it increases} from our base temperature of 7.4 to 300~K, corresponding to a negative thermal expansion coefficient along the $c$-axis over the full $T$ range of the measurement. In contrast, the $a$~lattice parameter {\it increases} by $\approx 0.35$\% with increasing $T$ over the same $T$ interval which corresponds to the usual positive thermal expansion coefficient.  The volume thermal expansion in Fig.~\ref{Tdep}(b) initially decreases on heating above 8~K and then increases from $\approx 80$~K to room temperature.

Negative thermal expansion coefficients in the paramagnetic states of materials are unusual.  Indeed, the anisotropic thermal expansion coefficients above the coupled structural and magnetic transitions in crystals of the $A{\rm Fe_2As_2}$ high-$T_{\rm c}$ parent compounds \cite{Budko2010} and in Ba(Fe$_{1-x}$Co$_x)_2$As$_2$ crystals with $x = 0.038$ and~0.074 (Ref.~\onlinecite{Budko2009}) are all positive.  On the other hand, Ca$_{1-x}$La$_x{\rm Fe_2As_2}$ crystals with $x = 0.15$, 0.20 and~0.25 all show temperature regions where the $a$- and/or $c$-axis and volume expansion coefficients are negative.\cite{Rebello2012}  These latter behaviors are suggested by the authors to originate from proximity to a structural phase transition.

Our thermal expansion data in Fig.~\ref{Tdep} reveal no obvious evidence for a temperature-induced lattice distortion in $\rm{SrCo_2As_2}$.  We therefore specifically searched for an orthorhombic lattice distortion, using high-resolution single-crystal x-ray diffraction, that might be associated with the observed negative $c$-axis and volume thermal expansion coefficients.  However, no splitting or changes of peak shape were observed in the temperature range 10--300~K for the (2~0~8) or (1~1~8) reflections measured in the $(h~0~\ell)$ and $(h~h~\ell)$ scattering planes, respectively, demonstrating that no tetragonal to orthorhombic lattice distortion occurs over this $T$ range to within an estimated upper limit for a potential distortion of $(a-b)/(a+b)\leq 2.5\times10^{-4}$.

\section{\label{Sec:rho} Electrical Resistivity}

\begin{figure}[t]
\includegraphics[width=3.3in]{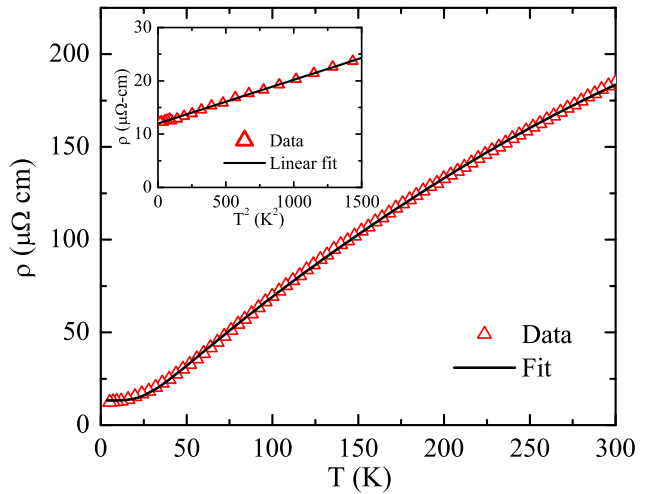}
\caption{(Color online) Electrical resistivity $\rho$ measured along the $ab$ plane versus temperature $T$  for $\rm{SrCo_2As_2}$. The solid curve is a fit by Eqs.~(\ref{Eqs:BGFit}). Inset: $\rho$ versus $T^2$, along with a linear fit to the data between 3.5 and 38~K (straight line).}
\label{fig:Figure_Res}
\end{figure}

Figure~\ref{fig:Figure_Res} shows the $ab$-plane $\rho(T)$ of $\rm{SrCo_2As_2}$. The data exhibit metallic behavior with a high residual resistivity ratio ${\rm RRR} \equiv \rho(300~{\rm K})/\rho(3~{\rm K}) = 15.3$, indicating good crystal quality.  At low temperatures ($3.5~{\rm K} \leq T \leq 38$~K) we find that $\rho(T)$ follows $\rho = \rho_0 + AT^2$ (inset of Fig.~\ref{fig:Figure_Res}) as expected for a Fermi liquid.\cite{Stewart2001} A fit by this equation gives $\rho_0 = 12.0~\mu\Omega$\,cm and $A=8.21(4)\,\times\,10^{-3}$~$\mu\Omega$~cm/K$^2$.

The data over the entire $T$~range could not be fitted well solely by the sum of a $T$-independent term $\rho_0$ and the Bloch-Gr\"uneisen prediction for carrier scattering by longitudinal lattice vibrations in the absence of umklapp scattering given by\cite{Goetsch2012, Blatt1968}
\bse
\label{Eqs:BGFit}
\begin{equation}
\rho_{\rm BG}(T)= 4 \mathcal{R} \left(\frac{T}{\Theta_{\rm R}}\right)^5 \int_0^{\Theta_{\rm R}/T}\frac{x^5}{(e^x-1)(1-e^{-x})}\,dx, 
\label{eq:Bloch-Gruneisen}
\end{equation}
where $\Theta_{\rm R}$ is the Debye temperature obtained from fitting resistivity measurements and $\mathcal{R}$ is a material-dependent scale factor that is independent of $T$\@.  Equation~(\ref{eq:Bloch-Gruneisen}) yields
\be
\rho(T=\Theta_{\rm R}) \approx 0.946\,464\,{\cal R}.
\label{Eq:CalR}
\ee
At high temperatures $T\gtrsim \Theta_{\rm R}$, the BG model yields a $\rho_{\rm BG}(T)$ that is proportional to~$T$, whereas the experimental data in Fig.~\ref{fig:Figure_Res} show negative curvature at high~$T$.  Therefore, we also used a Mott interband $sd$ scattering term $-D T^3$ (Refs.~\onlinecite{Mott1936, Giovannini1998}) to fit the data according to 
\begin{equation}
\rho(T) = \rho_0 + \rho_{\rm {BG}}(T) -D T^3.
\label{eq:BG_fit}
\end{equation}
This fit function is not accurate at low $T$ because the data below 38~K follow a $T^2$ dependence as shown in the inset of Fig.~\ref{fig:Figure_Res} and discussed above, whereas $\rho_{\rm BG}(T)$ has a $T^5$ dependence at low $T$\@.  Furthermore, we used the high-accuracy analytic dimensionless Pad\'e approximant $\rho_{\rm n}(T/\Theta_{\rm R})$ in Ref.~\onlinecite{Goetsch2012} in place of Eq.~(\ref{eq:Bloch-Gruneisen}) to facilitate the least-squares fit to the data, where
\be
\rho_{\rm {BG}}(T) = \rho(\Theta_{\rm R})\rho_{\rm n}(T/\Theta_{\rm R}).
\ee
\ese

A good fit of the experimental $\rho(T)$ data in Fig.~\ref{fig:Figure_Res} by Eqs.~(\ref{Eqs:BGFit}) was obtained (black curve in Fig.~\ref{fig:Figure_Res}), where the fitted parameters are $\rho_0 = 13.3(1)\,\mu\Omega$\,cm,  $\rho(\Theta_{\rm R}) = 104(1)\,\mu\Omega\,$cm, $\Theta_{\rm R} = 167(1)$~K and $D = 9.1(1)\times 10^{-7}$~$\mu\Omega$\,cm\,K$^{-3}$. From the value of $\rho(\Theta_{\rm R})$ and Eq.~(\ref{Eq:CalR}) we obtain ${\cal R} = 110\,\mu\Omega$\,cm.  A significant contribution from the Mott $sd$ interband scattering is expected in materials where narrow $d$ bands appear at the Fermi level, for example in elemental Pd metal.\cite{Mott1936} Since the ARPES data and the band structure calculations presented below show that an extremely narrow $d$ band is present at $E_{\rm F}$ in $\rm{SrCo_2As_2}$, a substantial contribution from Mott $sd$ interband scattering is not surprising.  

\section{\label{Sec:ARPES} ARPES}

\begin{figure*}
\includegraphics[width=6.in]{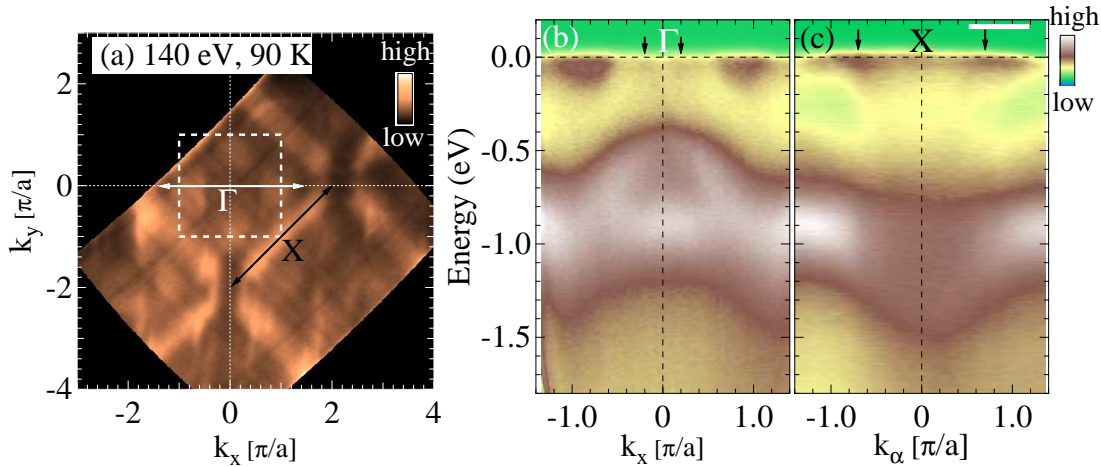}\vspace{-0.2in}
\caption{(Color online) (a) The Fermi surfaces of SrCo$_2$As$_2$ measured with 140~eV photon energy and at 90~K sample temperature. The photoemission intensity maps are obtained by integrating over an energy window of $\pm$10~meV with respect to the Fermi level. The outline of the first Brillouin zone (BZ) is shown by the white dashed square. (b) Band dispersion data plotted through the center [$\Gamma$-point, horizontal white line in (a)] of the BZ from ${\bf k} = (\sim-\pi/a,0)$ to $(\sim\pi/a,0)$ and (c) through the corner [X-point $(\pi/a,-\pi/a)$, black line in (a)] of the BZ described by ${\bf k} = (\frac{\pi}{a},-\frac{\pi}{a}) +{\bf k}_\alpha$, where ${\bf k}_\alpha = (k_\alpha,k_\alpha)$ and $k_\alpha$ ranges from zero to  $\approx \pm1.2\,\pi/a$.  The black vertical arrows at the tops of~(b) and~(c) denote the Fermi wave vector positions for the electron Fermi surface pockets centered at $\Gamma$ and X, respectively.}
\label{Fig:ARPES}
\end{figure*}

\begin{figure}
\includegraphics[width=2.8in]{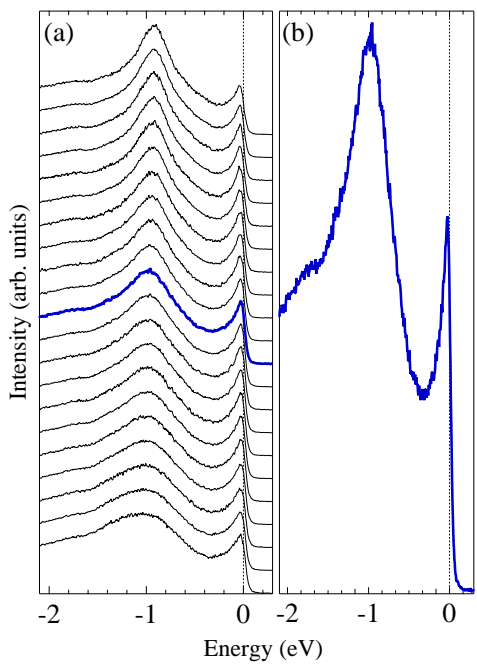}
\caption{(Color online) (a) Energy distribution curves (EDCs) at fixed wave vector as the wave vector is scanned through the $\Gamma$ point along the white bar at the top of Fig.~\ref{Fig:ARPES}(c), which is from $k_\alpha = 0.52\,\pi/a$ to~$1.19\,\pi/a$.  The EDC highlighted in bold blue is at $k_\alpha = 0.85\,\pi/a$.  The Fermi energy $E_{\rm F}$ is at the zero of energy on the horizontal scale.  (b)  Enlargement of the EDC at $k_\alpha = 0.85\,\pi/a$ that is plotted in bold blue in (a).  The sharp quasiparticle peak near $E_{\rm F}$ arises from a flat electron band with a sharp peak in the density of states near $E_{\rm F}$.}
\label{Fig:Fig2_ARPES_SrCo2As2}
\end{figure}

We studied the electronic properties of a SrCo$_2$As$_2$ single crystal using high-resolution ARPES, which can directly probe the low-energy electronic excitations as a function of binding energy and momentum. Figure~\ref{Fig:ARPES}(a) shows the photoemission intensity map, which is obtained by integrating over an energy window of $\pm$10~meV with respect to the Fermi energy $E_{\rm F}$, measured with a photon energy of 140~eV and sample temperature of 90~K\@.  We clearly see four patches of intensity arranged around the center ($\Gamma$ point) of the Brillouin zone (BZ), which are located at $\pm\pi/a$ along the $k_x$ and $k_y$ directions.  More interesting is the observation of a large {\bf k}-independent intensity near $E_{\rm F}$ along the $[\pm1,\pm1]$ directions passing through the corners (X~points) of the BZ\@.  This indicates the presence of a flat band near $E_{\rm F}$ that extends over a large fraction of the second BZ\@.  Our LDA band structure calculations in Fig.~\ref{Fig:FIGBAND}(c) below confirm the presence of a flat $d_{x^2-y^2}$ band situated at or very close to $E_{\rm F}$ that passes through the X~points in the $[\pm1,\pm1]$ directions, where here the $x$ and $y$ axes are along the $a$ and $b$ axes of the Co square lattice, respectively.   A wide flat band just above $E_{\rm F}$ with an associated very sharp peak in the density of states was also present in previous LDA band calculations for KCo$_2$As$_2$.\cite{Singh2009}

Energy distribution curves (EDCs) for fixed momenta moving along the white line at the top of Fig.~\ref{Fig:ARPES}(c) are shown in Fig.~\ref{Fig:Fig2_ARPES_SrCo2As2}(a), and an enlargement of the EDC in bold blue in Fig.~\ref{Fig:Fig2_ARPES_SrCo2As2}(a) is plotted in Fig.~\ref{Fig:Fig2_ARPES_SrCo2As2}(b).  This shows in more detail a large and sharp peak in the quasiparticle density of states at $E_{\rm F}$ as expected from the flat band near $E_{\rm F}$ and associated sharp peak in the density of states identified in the band structure calculations.  We speculate that the anomalous $c$-axis thermal expansion behavior and perhaps some of the magnetic properties we observe for SrCo$_2$As$_2$ are driven by the presence of this flat band at or near $E_{\rm F}$\@.

In general, this Fermi surface (FS) is very different from those of other related parent compounds of the 122 family like BaFe$_2$As$_2$ and SrFe$_2$As$_2$, where the hole FSs at $\Gamma$ and the electron FSs at~X are roughly circular in shape and similar in size.\cite{Kondo2010} To gain more insight about the character of the FS pockets, in Figs.~\ref{Fig:ARPES}(b,c) we show the energy-momentum ARPES intensity maps along two different cuts, marked by the white [across the $\Gamma$~(0,0) point]  and black [across the X~($\pi/a, -\pi/a$) point] bars in Fig.~\ref{Fig:ARPES}(a). In Fig.~\ref{Fig:ARPES}(b), one can observe a rather weak but discernible electron-like FS pocket at the $\Gamma$-point and two additional small electron pockets at  $\pm\pi/a$. A general observation in Ba(Fe$_{1-x}$Co$_x)_2$As$_2$ is that with increasing $x$ (electron doping), the central hole pockets at $\Gamma$ become electron-like, while the X~pockets remain electron-like and keep expanding on the overdoped side of the phase diagram (see, e.g., the ARPES data for the $x = 0.42$ sample in Ref.~\onlinecite{Liu2011}).  An electron pocket at the center of the BZ [Fig.~\ref{Fig:ARPES}(b)] for SrCo$_2$As$_2$ is expected and is in overall agreement with our previous study of Ba(Fe$_{1-x}$Co$_x$)$_2$As$_2$.\cite{Liu2011} The pocket at the X-point is also electron-like, but much larger in size [Fig.~\ref{Fig:ARPES}(c)], so there is no obvious Fermi surface nesting in SrCo$_2$As$_2$.

We extracted the Fermi momenta of the electron pockets from the peaks in the momentum distribution curves at $E_{\rm F}$ (not shown), and were found to be about $\pm 0.2$~\AA$^{-1}$ [vertical black arrows at the top of Fig.~\ref{Fig:ARPES}(b)] and $\pm 0.7$~\AA$^{-1}$ [vertical black arrows at the top of Fig.~\ref{Fig:ARPES}(c)] centered at the $\Gamma$ and X~points, respectively.  Figure~\ref{Fig:ARPES}(b) also shows that the top of the hole band at the $\Gamma$ point has moved from above $E_{\rm F}$ for the $A{\rm Fe_2As_2}$ compounds to about 400~meV below $E_{\rm F}$ due to Co contributing one conduction electron more than Fe.  This leads to the disappearance of the hole-like Fermi surface around the $\Gamma$-point as discussed above,\cite{Liu2011} and is consistent with our recent ARPES study of BaCo$_2$As$_2$.\cite{Dhaka2013}  In particular, the energies of the top and bottom of the valence band at the $\Gamma$ and~X points, the size of the electron pockets surrounding the $\Gamma$ point, and the location and extent of the flat band at the X~point are very similar for both materials.

\section{\label{Sec:BandTheory} Band Structure Calculations}

\begin{figure}\vspace{-0.1in}
\includegraphics[width=3.3in]{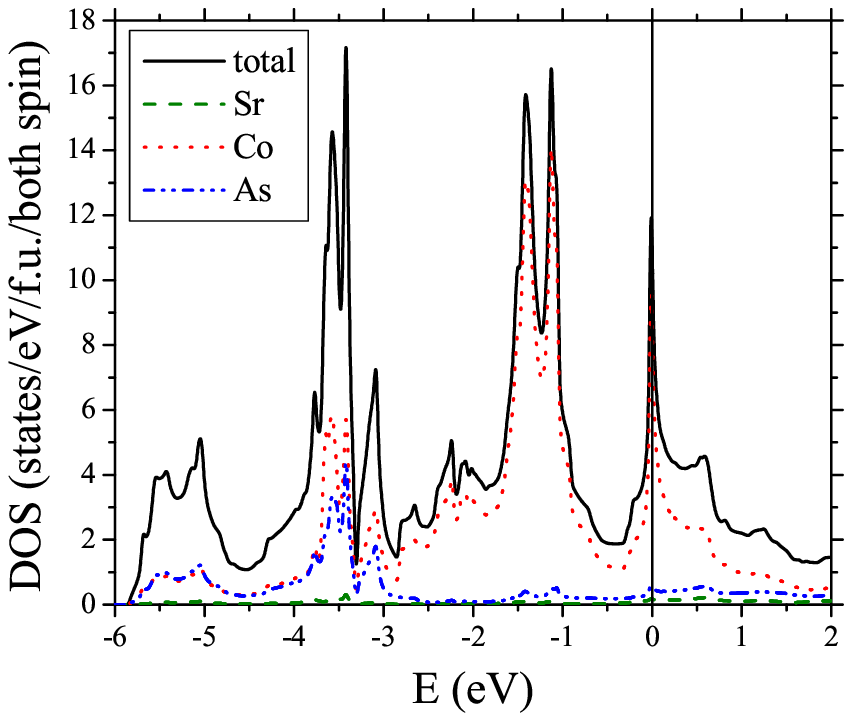}\vspace{-0.4in}
\includegraphics[width=3.3in]{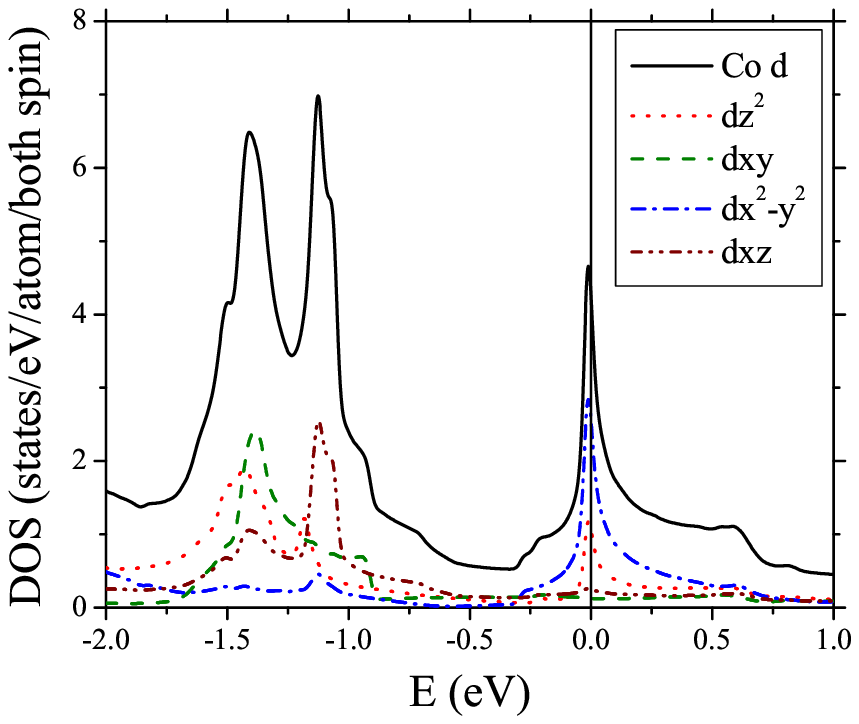}\vspace{-0.2in}
\caption{(Color online) Top panel: Total and atom-decomposed density of states (DOS) of $\rm{SrCo_2As_2}$ versus energy $E$\@.  Bottom panel: Orbital-decomposed contributions of the Co $3d$ atoms to the DOS\@.  The peak in the total DOS near $E_{\rm F}=0$ is mostly contributed by the Co atoms.  The DOS shows three peaks in the energy range shown. The peak near $E_{\rm F}$ is mainly from the $d_{x^2-y^2}$ flat-band component, and the second peak near 1.2~eV below $E_{\rm F}$ is from the $d_{xz},\ d_{yz}$ components which are degenerate in tetragonal symmetry.  The third peak near 1.5~eV below $E_{\rm F}$ is from the $d_{xy}$ and $d_{z^2}$ components.  The ordinate scales of the DOS are different in the two panels because the top panel is for all atoms whereas the bottom panel is for the Co atoms separately.}
\label{Fig:DOS}
\end{figure}

The top and bottom panels of Fig.~\ref{Fig:DOS} are atom- and orbital-decomposed densities of states (DOS) which were calculated with 1631 {\bf k}~points in the irreducible BZ\@. The top panel shows strongly hybridized bonding states between Co and As atoms in the energy range from 3 to 6~eV below $E_{\rm F}$.  There are also antibonding empty states but they are located outside of the energy range shown.  It also shows, near $E_{\rm F}$, a strong peak which is mostly of Co $d$~character.  It is well known that a system that has a high DOS or peak at $E_{\rm F}$ is unstable and has a tendency to split the peak (lower the total energy) via physical transformations such as magnetic or structural transitions.\cite{Mohn2006}  Therefore, it might be expected that ${\rm SrCo_2As_2}$ also shows peculiar temperature- or pressure-dependent properties, which depend on the precise peak position.

The bottom panel of Fig.~\ref{Fig:DOS} shows the orbital-decomposed Co atom DOS in a small energy range (from 2~eV below to 1~eV above $E_{\rm F}$) to emphasize the orbital contributions to the peak near $E_{\rm F}$.  The $d_{x^2-y^2}$ orbital points towards neighboring Co atoms in the $xy$-plane and is a dominant contributor, where the $x$ and $y$ axes are defined here to be along the $a$ and $b$ axes of the Co square lattice, respectively.  It suggests a possibility to control the peak size or position by Co--Co distance modification.

Thus the very sharp peak in the DOS near $E_{\rm F}$ in Fig.~\ref{Fig:DOS}(a) arises from a Co $d_{x^2-y^2}$ flat band extending over a large fraction of the second BZ\@.  The band structure density of states at $E_{\rm F}$ is
\be
{\cal D}^{\rm band}(E_{\rm F}) = 11.04~{\rm states/(eV\,f.u.)\ both\ spins}.  
\label{Eq:DEFband}
\ee

\begin{figure}
\includegraphics[width=3.4in]{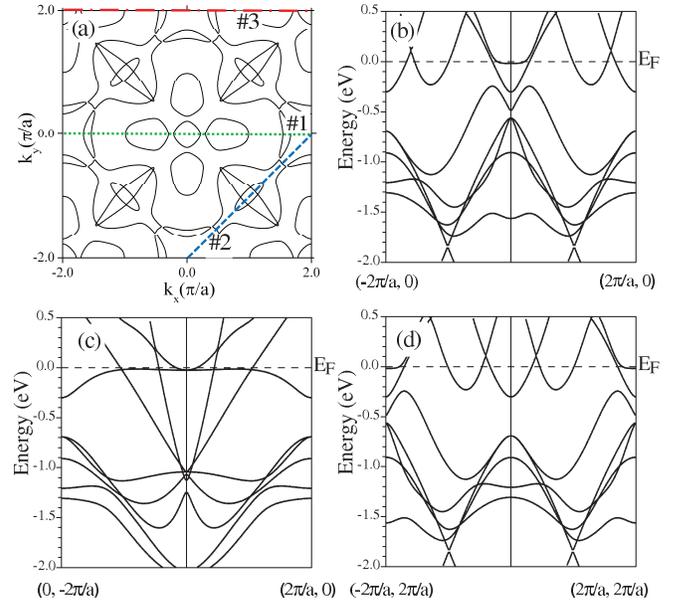}
\caption{(Color online) Plots of (a) the Fermi surface and (b--d) the band structure of ${\rm SrCo_2As_2}$ calculated at $k_z=1.88 (\pi/c)$. The lines (\#1\,--\,\#3) in~(a) are cutting lines for band structure plot directions and correspond to panels~(b)--(d), respectively.  The lattice parameters $a$ and $c$ are those of the body-centered tetragonal unit cell of ${\rm SrCo_2As_2}$ in Sec.~\ref{Sec:ExpDetails}.}
\label{Fig:FIGBAND}
\end{figure}

Figure~\ref{Fig:FIGBAND} shows a Fermi surface (FS) plot and band structures for three different cuts in {\bf k}-space. For the Fermi surface calculation, we divided the $-2\pi/a < k_x,\ k_y < 2\pi/a$ range of $k_x,~k_y$ planes with different $k_z$ values into a $200\times200$ mesh that gives 40401~${\bf k}$~points.  Figure~\ref{Fig:FIGBAND}(a) shows the result with $k_z = 1.88 (\pi/c)$ and also shows three cutting lines for band structure calculations.  The cutting line \#1 is from $(-2\pi/a,0,z)$ to $(2\pi/a,0,z)$, \#2 is from $(0, -2\pi/a,z)$ to $(2\pi/a,0,z)$ and cutting line \#3 is from $(-2\pi/a, 2\pi/a,z)$ to $(2\pi/a, 2\pi/a,z)$, with $z=1.88(\pi/c)$. These cutting lines \#1\,--\,\#3 correspond to the band structure plots Figs.~\ref{Fig:FIGBAND}(b)--\ref{Fig:FIGBAND}(d), respectively. Figure~\ref{Fig:FIGBAND}(c) shows the $d_{x^2-y^2}$ flat band that is largely responsible for the sharp high peak in the DOS near $E_{\rm F}$.\@  

\section{\label{Sec:Cp} Heat Capacity}

\begin{figure}
\includegraphics[width=3.3in]{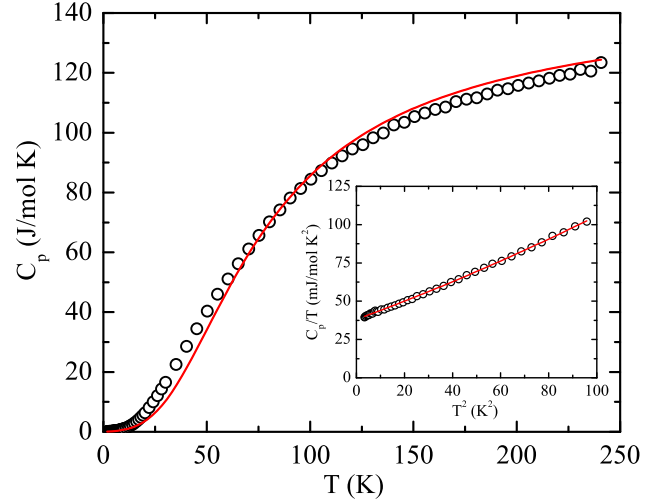}
\caption{(Color online) Heat capacity $C_{\rm p}$ versus temperature $T$ for $\rm{SrCo_2As_2}$. The red curve is a fit by Eqs.~(\ref{Eqs:CpFit}). Inset: $C_{\rm p}/T$ versus~$T^2$. The curve is a fit by Eq.~(\ref{Eq:Cp/Tfit}).}
\label{fig:Figure_HC}
\end{figure}

The $C_{\rm p}$ of $\rm{SrCo_2As_2}$ is plotted versus~$T$ in Fig.~\ref{fig:Figure_HC}. The value of $C_{\rm p}$ at 240~K is 124~J/mol~K which is close to the Dulong-Petit high-$T$ limit of the lattice heat capacity at constant volume given by $C_{\rm V} = 15R \approx 124.7$~J/mol~K, where $R$ is the molar gas constant.  The inset in Fig.~\ref{fig:Figure_HC} shows a conventional $C_{\rm p}/T$ versus $T^2$ plot for $T \leq 10$~K where the data are fitted by the expression
\be
C_{\rm p}/T =  \gamma + \beta T^2 + \delta T^4
\label{Eq:Cp/Tfit}
\ee
which is the sum of electronic $(\gamma T)$ and low-$T$ lattice ($\beta T^3 +\delta T^5)$ heat capacity contributions. The fitted values of the Sommerfeld coefficient $\gamma $ and the lattice coefficients $\beta$ and $\delta$ are listed in Table~\ref{Table:Parameters}. The value of $\gamma = 37.8(1)~{\rm mJ/mol\,K^2}$ is very large compared to those of the parent iron arsenide compounds $A\rm{Fe_2}As_2$ ($A$ = Ca, Sr, Ba) in the stripe AFM state.\cite{Ronning2008, Chen2008, Rotundu2010}  Additional $C_{\rm p}(T)$ data at $H=3$~T down to 1.3~K revealed no new features (not shown).

The density of states at the Fermi energy for both spin directions ${\cal D}(E_{\rm F})$, including many-body enhancement effects due to the electron-phonon and electron-electron interactions, is calculated from
\be
\gamma = \frac{\pi^2k_{\rm B}^2}{3}{\cal D}^\gamma(E_{\rm F}).
\ee 
Using the above value of $\gamma$ yields
\be
{\cal D}^\gamma(E_{\rm F}) = 16.0(3)~{\rm states/eV\,f.u.}
\label{Eq:DEFgamma}
\ee
for both spin directions.  This value is 1.45 times the band theory value of 11.04~states/(eV\,f.u.) (Table~\ref{Table:Parameters}).  Writing $\gamma = \gamma_{\rm band}(1+\lambda_{\rm el-ph})m^\ast/m_{\rm band}$, where $\gamma_{\rm band}$ and $m_{\rm band}$ are the band theory values of $\gamma$ and the current carrier effective mass, $\lambda_{\rm el-ph}$ is the electron-phonon coupling constant and $m^\ast/m_{\rm band}$ is the electronic many-body conduction carrier mass enhancement over the band theory value, gives 
\be
(1+\lambda_{\rm el-ph})m^\ast/m_{\rm band} \approx 1.45.
\label{Eq:lambdaeph}
\ee
A similar analysis of $\gamma$ obtained at low~$T$ has been done for paramagnetic $\rm{BaCo_2As_2}$, where the authors find\cite{Sefat2009}
\bse
\bea
\gamma &=& 41.0~\rm{mJ/mol~K^2},\\*
{\cal D}^\gamma(E_{\rm F}) &=& 17.4~{\rm states/eV\,f.u.}, \label{Eq:DEFgammaBaCo2As2}\\*
&&\hspace{-0.6in} (1+\lambda_{\rm el-ph})m^\ast/m_{\rm band} \approx 2.0.\label{Eq:lambdm*BaCo2As2}
\eea
\ese
The $\gamma$ and ${\cal D}^\gamma(E_{\rm F})$ values of $\rm{SrCo_2As_2}$ and $\rm{BaCo_2As_2}$, respectively, are seen to be nearly the same.  The difference between the combined electron-phonon and electronic mass enhancements in Eqs.~(\ref{Eq:lambdaeph}) and~(\ref{Eq:lambdm*BaCo2As2}) arises mainly from the difference in the band structure densities of states of the two compounds.

The Debye temperature $\Theta_{\rm D}$ is obtained from the value of $\beta$ according to
\be
\Theta_{\rm D} = \left(\frac{12\pi^4Rn}{5\beta}\right)^{1/3},
\ee
where $\beta$ is normalized to a mole of formula units (f.u.)\ and $n=5$ is the number of atoms per f.u.  Thus we obtain $\Theta_{\rm D} = 251(1)$~K\@.  This value is much larger than the value $\Theta_{\rm R} = 167(1)$~K obtained above in Sec.~\ref{Sec:rho} from the Bloch-Gr\"uneisen fit to $\rho(T)$.  This discrepancy may be associated with the unusual negative thermal expansion characteristics discussed in Sec.~\ref{Sec:ThermalExpansion} above and/or with $T$-dependent spin fluctuation effects discussed in Sec.~\ref{Sec:NMR} below and associated magnetoelastic coupling effects. 

The Kadowaki-Woods (KW) ratio is $R_{\rm KW} = A/\gamma^2 = 5.7\,\times\,10^{-3}$~m$\Omega$\,cm\,mol$^2$\,J$^{-2}$\,K$^2$, similar to those observed in strongly-correlated electron metals and heavy fermion compounds,\cite{Jacko2009} thus suggesting the presence of strong electron correlations in $\rm{SrCo_2As_2}$.

The $C_{\rm p}(T)$ data in the entire $T$~range of measurement were fitted by 
\bse
\label{Eqs:CpFit}
\begin{equation}
C_{\rm p}(T) = \gamma T + nC_{\rm{V\,Debye}}(T) ,
\label{Eq:DebyeHC}
\end{equation}
where $\gamma$ was fixed at the above-determined value 37.8~mJ/mol~K$^2$ and $n=5$ is the number of atoms/f.u.  The function $C_{\rm{V\,Debye}}(T)$ is the Debye lattice heat capacity at constant volume per mole of atoms due to acoustic phonons given by\cite{Gopal1966} 
\begin{equation}
C_{\rm{V\,Debye}}(T) = 9 R \left( \frac{T}{\Theta_{\rm{D}}} \right)^3 {\int_0^{\Theta_{\rm{D}}/T} \frac{x^4 e^x}{(e^x-1)^2}\,dx},
\label{eq:Debye_HC}
\end{equation}
\ese
where $\Theta_{\rm D}$ is the Debye temperature.  A high-accuracy analytic Pad\'e approximant\cite{Goetsch2012} for the Debye function in Eq.~(\ref{eq:Debye_HC}) was used for fitting the $C_{\rm p}(T)$ data which greatly facilitates the fit. The $\Theta_{\rm D}$ derived from the least-squares fit of the data over the full $T$ range in Fig.~\ref{fig:Figure_HC} by Eq.~(\ref{Eq:DebyeHC}) is 304(3)~K as listed in Table~\ref{Table:Parameters}.  This value of $\Theta_{\rm D}$ is significantly larger than the value of 251(1)~K determined above from fitting the $C_{\rm p}$ data at low~$T<10$~K by Eq.~(\ref{Eq:Cp/Tfit}). Furthermore, the goodness of fit obtained is poor, evident from the large systematic deviations of the fit from the data in Fig.~\ref{fig:Figure_HC} compared to the much better fits we recently obtained for other 122-type compounds.\cite{Goetsch2012, Anand2013} This unusually large deviation between the two $\Theta_{\rm D}$ values might be related to the negative thermal expansion reported in Sec.~\ref{Sec:ThermalExpansion} and/or magnetoelastic coupling effects arising from strongly $T$-dependent spin fluctuations. 

The parameters of the above fits and derived quantities are summarized in Table~\ref{Table:Parameters}.

\begin{table}
\caption{Parameters derived from $C_{\rm p}(T)$, $\rho(T)$ and $\chi$ data of $A\rm{Co_2As_2}$ ($A$ = Sr, Ba). The listed parameters are the Sommerfeld coefficient $\gamma$ and the coefficients $\beta$ and $\delta$ of the lattice heat capacity at low $T$ in Eq.~(\ref{Eq:Cp/Tfit}), density of states at the Fermi energy obtained from $T\to0$  heat capacity ${\cal D}^\gamma(E_{\rm F})$ and magnetic susceptibility ${\cal D}^\chi(E_{\rm F})$ measurements, respectively, Wilson ratio $R_{\rm W} = {\cal D}^\chi(E_{\rm F})/{\cal D}^\gamma(E_{\rm F})$, density of states at the Fermi energy obtained from band structure calculations ${\cal D}^{\rm band}(E_{\rm F})$, Debye temperatures $\Theta_{\rm D}$ and $\Theta_{\rm R}$ determined from heat capacity and resistivity measurements, respectively, coefficient $A$ of the $T^2$ term in $\rho$ at low~$T$, the Kadowaki-Woods ratio $R_{\rm KW} = A/\gamma^2$, the electron-phonon coupling constant $\lambda_{\rm el-ph}$ and the electronic many-body mass enhancement $m^\ast/m_{\rm band}$. The $\beta$ value for $\rm{BaCo_2As_2}$ was calculated by digitizing the $C_{\rm p}/T$ versus $T^{2}$ data in Ref.~\onlinecite{Sefat2009}.}
\label{Table:Parameters}
 \begin{ruledtabular}
		\begin{tabular}{l c c  }
		 Parameter                                        & $\rm{SrCo_2As_2}$     & $\rm {BaCo_2As_2}$ \\
		                                                & (This work) & (Refs.~\onlinecite{Sefat2009} and \onlinecite{SafatError})\\
	 \hline
			$\gamma$ (mJ/mol~K$^2$)                                     & 37.8(1)                & 41.0 \\
			$\beta$ (mJ/mol~K$^4$)                                      & 0.611(7)               & 0.47 \\
			$\delta\ (\mu$J/mol~K$^6$)                                   & 1.01(9)                &  \\
			${\cal D}^\gamma(E_{\rm F})$\footnotemark[1]   			& 16.0(3)                & 17.4 \\
			${\cal D}^\chi(E_{\rm F})$\footnotemark[1]     			& 54                	& 90 \\
			$R_{\rm W}$										& 3.4				& 5.2 \\
			${\cal D}^{\rm band}(E_{\rm F})$\footnotemark[1]    		& 11.04                	& 8.5 \\
			$\Theta_{\rm D}$ (K) (Low $T$)                     		& 251(1)                 & 274\\
			$\Theta_{\rm D}$ (K) (All $T$)                			& 304(3)                 & \\
			$\Theta_{\rm R}$ (K) (All $T$)                			& 167(1)                 & \\
			$A$ ($10^{-3}$\,$\mu\Omega$\,cm/K$^2$)                      & 7.92(3)                & 2.2 \\
			${R_{\rm KW}}$\footnotemark[2]                                 & 5.7                   & 1.3 \\
			$(1+\lambda_{\rm el-ph})m^\ast/m_{\rm band}$     			& 1.45                   & 2.0 \\	
		\end{tabular}
\end{ruledtabular}
\footnotetext[1]{The units are ${\rm states/(eV\,f.u.)}$ for both spin directions.}
\footnotetext[2]{The units are ${\rm 10^{-3}~m\Omega\,cm\,mol^2\,J^{-2}\,K^2}$.}
\end{table}

\section{\label{Sec:M_chi} Magnetization and Magnetic Susceptibility}

\begin{figure}[h]
\includegraphics[width=3.3in]{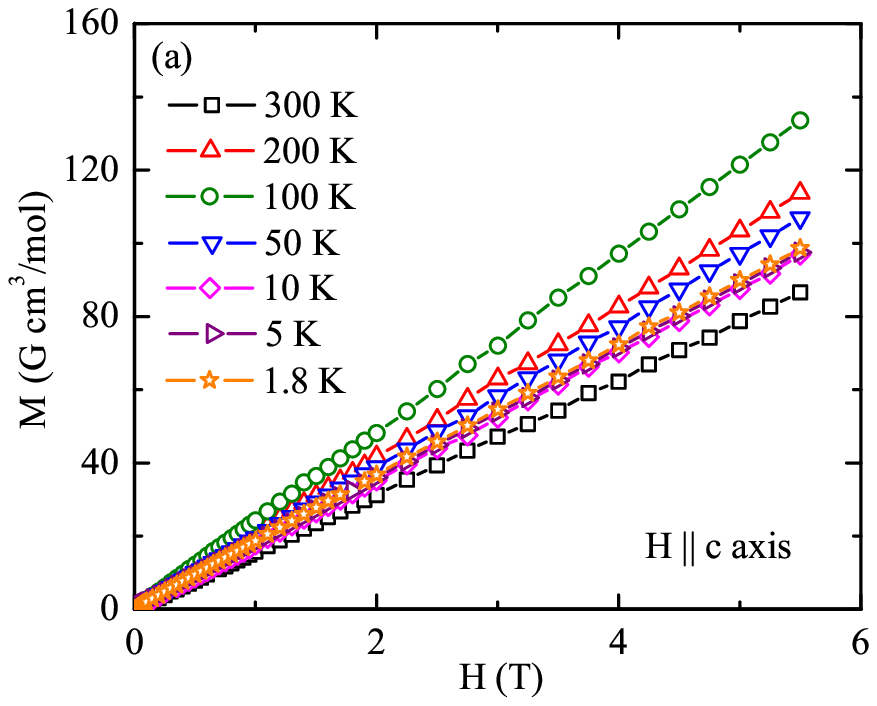}
\includegraphics[width=3.3in]{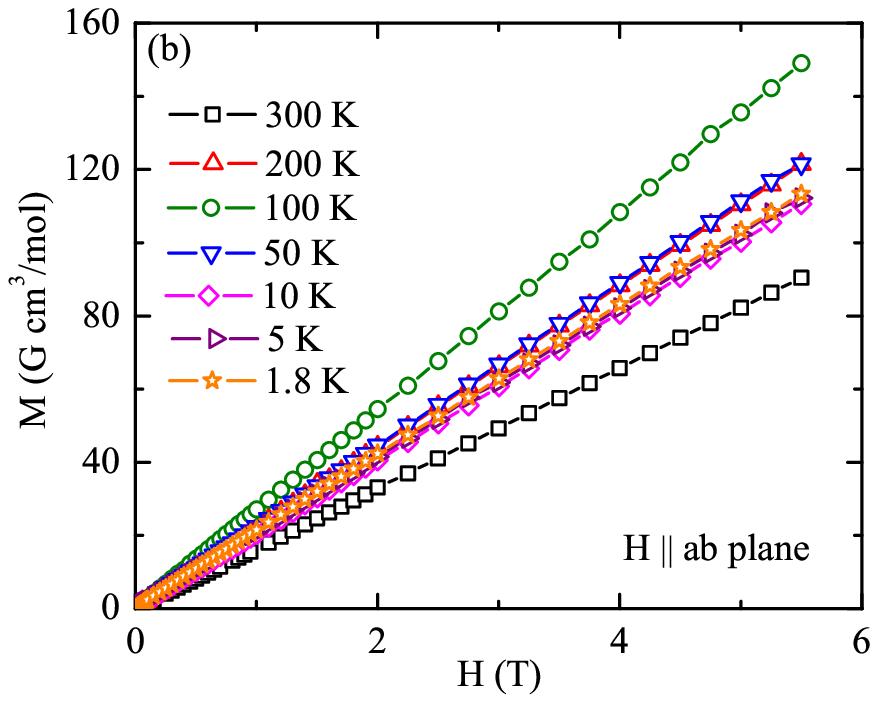}
\caption{(Color online) Isothermal magnetization $M$ of $\rm{SrCo_2As_2}$ versus applied magnetic field $H$ at the indicated temperatures for $H$ parallel to (a) the $c$-axis and (b) the $ab$~plane.}
\label{fig:Figure_MH}
\end{figure} 

\begin{figure}
\includegraphics[width=3.3in]{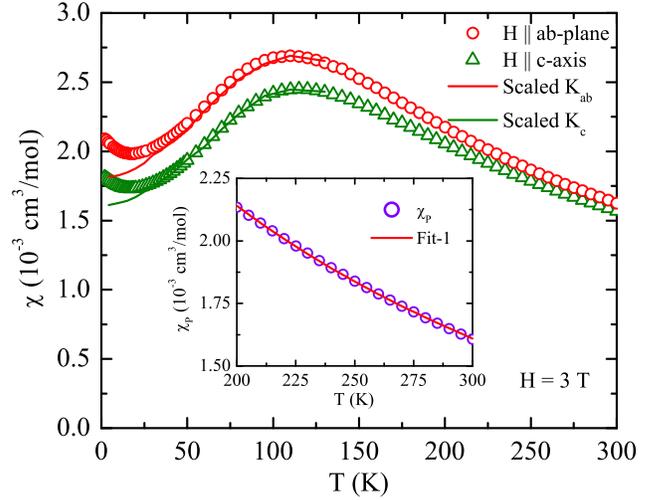}
\caption{(Color online) Anisotropic magnetic susceptibility $\chi \equiv M/H$ versus temperature $T$ (where $M$ is magnetization and $H$ is applied magnetic field) of $\rm{SrCo_2As_2}$ measured at $H=3$~T. From our NMR Knight shift measurements in Fig.~\ref{fig:K-chi}(a) below, the upturns in $\chi(T)$ below about 20~K are not intrinsic.  The solid curves are these Knight shift data scaled onto the low-$T$ $\chi(T)$ data, allowing estimates of the intrinsic $\chi(T\to0)$ values to be obtained as given in Eqs.~(\ref{Eqs:IntrinsicChiSr}).  Inset: Fit of the powder-averaged susceptibility $\chi_{\rm P}$ versus $T$ for $200~{\rm K} \leq T \leq 300$~K by the Curie-Weiss law in Eq.~(\ref{Eq:CWFit}) with $\chi_0 \equiv 0$ (Fit~1 in Table~\ref{Table:CWFitPars}).  Fits~1 and~2 are indistinguishable on the scale of the inset.}
\label{fig:Figure_Mag}
\end{figure}

The isothermal magnetization $M$ of a $\rm{SrCo_2As_2}$ single crystal versus $H$ for $H\parallel c$ and $H\parallel ab$-plane are shown in Figs.~\ref{fig:Figure_MH}(a) and \ref{fig:Figure_MH}(b), respectively. The $M$ is proportional to $H$ for the entire $T$ and $H$ ranges of the measurements for both field directions, showing that the crystal does not contain detectable ferromagnetic (FM) or saturable PM impurities. Therefore the susceptibility can be defined at arbitrary field $H$ as $\chi=M(H)/H$.

The anisotropic magnetic susceptibilities $\chi_{ab}$ and $\chi_{c}$ of $\rm{SrCo_2As_2}$ in $H = 3$~T are plotted versus $T$ in Fig.~\ref{fig:Figure_Mag}, where $\chi_c\equiv\chi(H\parallel c)$ and $\chi_{ab}\equiv\chi(H\parallel ab$-plane). The $\chi(T)$ data for both field directions exhibit broad maxima at $\approx 115$~K followed by shallow minima at $\approx 20$~K and smooth low-$T$ upturns below 20~K, with no evidence for long-range magnetic ordering.    A very similar broad maximum in $\chi$ at about the same temperature was observed for isostructural ${\rm KFe_2As_2}$.\cite{Hardy2013} The maximum in $\chi(T)$ for that compound was attributed to a crossover from a coherent to an incoherent Fermi liquid state with increasing~$T$\@.\cite{Hardy2013}

The $T$ dependences of the $^{75}$As NMR Knight shift $K(T)$ data in Fig.~\ref{fig:K-chi}(a) below are in overall agreement with the $\chi(T)$ data in Fig.~\ref{fig:Figure_Mag}, except that the $K(T)$ data do not show the upturns seen in the $\chi(T)$ data below $\approx 20$~K\@.  This difference indicates that these upturns in $\chi(T)$ are not intrinsic and therefore likely arise from paramagnetic impurities or defects that are not saturable in our field range (see Fig.~\ref{fig:Figure_MH}).  The Knight shift data from Fig.~\ref{fig:K-chi}(a) are therefore scaled onto the respective low-$T$ $\chi(T)$ data in Fig.~\ref{fig:Figure_Mag} as shown by the solid curves, which represent the intrinsic $\chi(T)$ at low~$T$\@.  The intrinsic values for $\rm{SrCo_2As_2}$ at 4.3~K are found to be
\bse
\label{Eqs:IntrinsicChiSr}
\bea
\chi_{c}(0) &=& 1.61\times10^{-3}\ {\rm cm^3/mol},\\*
\chi_{ab}(0) &=& 1.82\times10^{-3}\ {\rm cm^3/mol},\\*
\chi_{\rm P}(0) &=& 1.75\times10^{-3}\ {\rm cm^3/mol},
\eea
where the powder-averaged susceptibility $\chi_{\rm P}$ is
\be
\chi_{\rm P} = \frac{1}{3}\chi_c + \frac{2}{3}\chi_{ab}.
\ee
\ese
The effective density of states at $E_{\rm F}$ for both spin directions ${\cal D}^\chi(E_{\rm F})$ obtained from magnetic susceptibility measurements is calculated from $\chi_{\rm P}(0)$ for $g=2$ using the expression
\be
\chi_{\rm P}(0) = \mu_{\rm B}^2{\cal D}^\chi(E_{\rm F}),
\label{Eq:chiDEF}
\ee
yielding a large density of states
\be
{\cal D}^\chi(E_{\rm F}) = 54\ {\rm states/(eV\,f.u.).}
\label{Eq:DEFchi}
\ee
The anisotropy in $\chi$ is small with $\chi_{ab}(0)$ slightly larger than $\chi_{c}(0)$. The sign of the anisotropy is opposite to that in the PM state of $\rm{CaCo_2As_2}$ (Refs.~\onlinecite{Cheng2012,Ying2012}) and in $\rm{BaCo_2As_2}$.\cite{Sefat2009, SafatError}  

For comparison, from the high-field slopes from 2 to 6.5~T of the $M(H)$ data for $\rm{BaCo_2As_2}$ at $T=1.8$~K in the inset of Fig.~1 of Ref.~\onlinecite{Sefat2009}, we obtain\cite{SafatError}
\bse
\label{Eqs:chiBaCo2As2}
\bea
\chi_c(0) &=& 3.28\times10^{-3}~{\rm cm^3/mol}\\*
\chi_{ab}(0) &=& 2.72\times10^{-3}~{\rm cm^3/mol}\ ({\rm BaCo_2As_2})\\*
\chi_{\rm P}(0) &=& 2.91\times10^{-3}~{\rm cm^3/mol} .
\eea
\ese
These values are $\sim70$\% larger than our respective values at 4.3~K for $\rm{SrCo_2As_2}$ in Eqs.~(\ref{Eqs:IntrinsicChiSr}).  The density of states calculated from $\chi_{\rm P}(0)$ for ${\rm BaCo_2As_2}$ using Eq.~(\ref{Eq:chiDEF}) is
\be
{\cal D}^\chi(E_{\rm F}) = 90\ {\rm states/(eV\,f.u.)\quad ({\rm BaCo_2As_2}).}
\label{Eq:DEFchiBaCo2As2}
\ee

\subsection{\label{Sec:Stoner} Stoner Enhancement and the Wilson Ratio:\\ A Conundrum}

In an itinerant picture, the Stoner-enhanced susceptibility at ${\bf q}=0$ and $T=0$ is\cite{Mizutani2003}
\be
\chi = \frac{\chi_0}{1-{\cal D}_{\rm band}(E_{\rm F})I/4} \equiv \frac{\chi_0}{1-F},
\label{Eq:StonerChi}
\ee
where $\chi_0$ is the conduction carrier spin susceptibility calculated from the band-structure density of states at $E_{\rm F}$ as in Eq.~(\ref{Eq:chiDEF}), $I\sim 0.8$~eV is the Stoner factor for Co (Ref.~\onlinecite{Sefat2009}) and the factor of 1/4 arises due to the presence of two Co atoms per f.u.\ and from the two spin directions counted in ${\cal D}(E_{\rm F})$.  Using ${\cal D}_{\rm band}(E_{\rm F}) = 11.04$~states/(eV\,f.u.) from Eq.~(\ref{Eq:DEFband}) and $I=0.8$~eV gives
\be
F = 2.2>1,
\ee
which from comparison with Eq.~(\ref{Eq:StonerChi}) indicates that ${\rm SrCo_2As_2}$ should have a FM ground state.  This inference is wrong because it strongly disagrees with our $M(H)$ and $\chi(T)$ data in Figs.~\ref{fig:Figure_MH} and~\ref{fig:Figure_Mag}, respectively.

The Wilson ratio $R_{\rm W}$ is defined as
\be
R_{\rm W} \equiv \frac{{\cal D}^\chi(E_{\rm F})}{{\cal D}^\gamma(E_{\rm F})}.
\label{Eq:RWDef}
\ee
Within an itinerant picture, if $R_{\rm W}\sim1$--2, $\chi$ and $\gamma$ are approximately equally enhanced above the band structure prediction, indicating that any enhancement of $\chi$ and $\gamma$ is due to a many-body enhancement of ${\cal D}(E_{\rm F})$.  On the other hand, if $R_{\rm W}\gg1$ or $R_{\rm W}\ll 1$, one would infer that strong FM or AFM conduction spin correlations are present, respectively, which change $\chi$ but not $\gamma$.\cite{Grimvall1976}  Using the values of ${\cal D}^\chi(E_{\rm F})$ and ${\cal D}^\gamma(E_{\rm F})$ in Eqs.~(\ref{Eq:DEFchi}) and~(\ref{Eq:DEFgamma}), respectively, Eq.~(\ref{Eq:RWDef}) gives
\be
R_{\rm W} = 3.5\quad {\rm for\ SrCo_2As_2}.
\label{Eq:RWSrCo2As2}
\ee
As discussed above, this large value of $R_{\rm W}$ indicates a significant FM enhancement of $\chi$, qualitatively consistent with the large value of the factor $F$ derived above.

These indications of dominant FM correlations between the itinerant current carriers in ${\rm SrCo_2As_2}$ are at odds with the broad maximum in $\chi(T)$ at about 115~K in Fig.~\ref{fig:Figure_Mag}, which instead suggests the dominance of low-dimensional AFM correlations,\cite{Johnston1997} and with the linear $M(H)$ behaviors in Fig.~\ref{fig:Figure_MH} that show no tendency towards saturation.  Indeed, inelastic neutron scattering measurements on this compound showed strong AFM fluctuations at the stripe AFM wave vector,\cite{Jayasekara2013} but preliminary searches for FM spin fluctuations in the first BZ yielded no detectable signal.  On the other hand, electronic structure calculations of the static $\chi({\bf q})$ showed enhancements at both the FM and stripe AFM wave vectors.\cite{Jayasekara2013}  Interestingly, enhancements of $\chi({\bf q})$ at the same two wave vectors are found from electronic structure calculations and $\chi$ measurements of the FeAs-based materials.\cite{Johnston2010, Mazin2008}

A similar analysis of $\chi$ and $C_{\rm p}$ data for isostructural ${\rm BaCo_2As_2}$ yielded a similar conclusion about the importance of FM itinerant spin correlations in this compound.\cite{Sefat2009}  Using the densities of states for ${\rm BaCo_2As_2}$ in Eqs.~(\ref{Eq:DEFgammaBaCo2As2}) and~(\ref{Eq:DEFchiBaCo2As2}), one obtains the value of the Wilson ratio from Eq.~(\ref{Eq:RWDef}) as
\be
R_{\rm W} = 5.2\quad {\rm for\ BaCo_2As_2}.
\ee
This value is larger than for ${\rm SrCo_2As_2}$ in Eq.~(\ref{Eq:RWSrCo2As2}), suggesting stronger FM correlations in this compound.  The authors of Ref.~\onlinecite{Sefat2009} suggested that the suppression of long-range FM order in ${\rm BaCo_2As_2}$ results from strong spin fluctuations associated with a nearby quantum critical point.

\subsection{Curie-Weiss Model}

\begin{table}
\caption{Parameters derived from fitting the powder-averaged magnetic susceptibility $\chi_{\rm P}(T)$ in the $T$ range $200\leq T \leq 300$~K  by the Curie-Weiss law~(\ref{Eq:CWFit}).  The error bars reflect systematic errors found by varying the temperature range of the fits.  The $g$-factor and effective moment $\mu_{\rm eff}$ are calculated using Eqs.~(\ref{Eq:CurieConst}) and~(\ref{Eq:mueff}), respectively, assuming spin $S=1/2$.}
\label{Table:CWFitPars}
\begin{ruledtabular}
\begin{tabular}{l c c  }
Parameter   & Fit~1 & Fit~2 \\
\hline
$\chi_0~(10^{-4}~{\rm cm^3/mol})$		&	$\equiv 0$	& $-4(2)$	\\
$C\ {\rm(cm^3\,K/mol) }$ & 0.65(1)  &  1.0(2)\\
$g$					& 1.86(2)	& 2.3(2)	\\
$\mu_{\rm eff}~(\mu_{\rm B}$/Co)	&  1.61(2)    & 	2.0(2)	\\
$\theta_{\rm p}$ (K) & $-102(7)$ & $-180(50)$ \\
\end{tabular}
\end{ruledtabular}
\end{table}

Above $\approx 200$~K, the $\chi(T)$ of $\rm{SrCo_2As_2}$ resembles a Curie-Weiss-like (CW) behavior
\be
\chi = \chi_0 + \frac{C}{T-\theta_{\rm p}}.
\label{Eq:CWFit}
\ee
The values of the Curie constant $C$ and Weiss temperature $\theta_{\rm p}$ obtained by fitting  $\chi_{\rm P}(T)$ between 200 and 300~K, assuming $\chi_0=0$ (Fit~1), are listed in Table~\ref{Table:CWFitPars}.  In a second fit, the $\chi_0$ value was allowed to vary, yielding a different set of fit parameters (Fit~2) in Table~\ref{Table:CWFitPars}.  The Fit~1 is shown as the red curve in the inset of Fig.~\ref{fig:Figure_Mag}; both fits to the data are equally good and cannot be distinguished on the scale of the inset.  In a local-moment model, the Curie constant per mole of spins is given by
\be
C = \frac{N_{\rm A}g^2S(S+1)\mu_{\rm B}^2}{3k_{\rm B}},
\label{Eq:CurieConst}
\ee
where $N_{\rm A}$ is Avogadro's number, $g$ is the spectroscopic splitting factor ($g$-factor), $S$ is the spin quantum number, $\mu_{\rm B}$ is the Bohr magneton and $k_{\rm B}$ is Boltzmann's constant. The corresponding effective magnetic moment is computed from 
\be
\mu_{\rm eff} = g\sqrt{S(S+1)}\,\mu_{\rm B}/{\rm Co}.
\label{Eq:mueff}
\ee
The calculated values of $g$ and $\mu_{\rm eff}$ for the two fits taking $S=1/2$ are given in Table~\ref{Table:CWFitPars}.  

For an interacting local moment Heisenberg model ${\cal H} = \sum_{\langle ij\rangle}J_{ij}{\bf S}_i\cdot{\bf S}_j$ for identical crystallographically equivalent spins, one has\cite{Johnston2012}
\be
\theta_{\rm p}= -\frac{S(S+1)}{3k_{\rm B}}\sum_jJ_{ij},
\ee
where the sum is over all interacting neighbors $j$ of a given spin~$i$.  Within the $J_1$-$J_2$ model for the square Co lattice where $J_1$ and $J_2$ are the four nearest- and four next-nearest-neighbor exchange interactions, respectively, and using $S=1/2$, one obtains
\be
\theta_{\rm p}= -\frac{J_1+J_2}{k_{\rm B}}.
\ee
Our $\theta_{\rm p}$ values then give
\bse
\bea
\frac{J_1+J_2}{k_{\rm B}} &=& 102(7)~{\rm K},\qquad {\rm (Fit~1)}\\*
\frac{J_1+J_2}{k_{\rm B}} &=& 180(50)~{\rm K},\hspace{0.22in} {\rm (Fit~2)}
\eea
\ese
which are strongly antiferromagnetic.

We emphasize that the values of $C,\ \mu_{\rm eff}, \theta_{\rm p}$ and \mbox{$J_1+J_2$} may be in error because the CW fit to the $\chi(T)$ data may not have been done at sufficiently high temperatures for the CW law to accurately describe the data, and/or because $\rm{SrCo_2As_2}$ may be an itinerant magnetic system instead of a local-moment one.

The results of $\chi(T)$ measurements on a polycrystalline $\rm{SrCo_2As_2}$ sample were reported earlier by Leithe-Jasper {\it et al.}\cite{Leithe-Jasper2008} The authors found that $\rm{SrCo_2As_2}$ is a CW paramagnet showing no magnetic ordering above 1.8~K with $C = 0.53~{\rm cm^3\,K/mol}$ and $\theta_{\rm p} = -29$~K\@. These $C$ and $|\theta_{\rm p}|$ values are both significantly smaller than our values in Table~\ref{Table:CWFitPars} for a single crystal, and these authors did not report observing the maximum in $\chi(T)$ that we observe at $\approx 115$~K in Fig.~\ref{fig:Figure_Mag} for both field orientations and therefore which also occurs for the polycrystalline average.

\subsection{\label{Sec:ICF} Interconfigurational Fluctuation (ICF) Model}

As discussed above, in an itinerant picture, one expects ${\rm SrCo_2As_2}$ and ${\rm BaCo_2As_2}$ to be ferromagnetic.  The observed  absence of ferromagnetism in these compounds suggests that strong spin fluctuation effects prevent the occurrence of long-range FM ordering, perhaps due to proximity to a quantum critical point.\cite{Sefat2009}  The observed broad maximum in $\chi(T)$ in Fig.~\ref{fig:Figure_Mag} and the large magnitude of $\theta_{\rm p}$ in Table~\ref{Table:CWFitPars} can then be analyzed within a phenomenological interconfigurational fluctuation (ICF) model where the average fluctuating valence of the magnetic ions in a material is $T$-dependent.\cite{Sales1975, Franz1980}  For this model, at high $T$, we find a CW-like $\chi(T)$ with a negative (AFM-like) Weiss temperature, as observed, which conflicts with the positive (FM-like) Weiss temperature expected if $\rm{SrCo_2As_2}$ were near a FM instability.  A many-body theory of intermediate-valence and interconfigurational fluctuation effects\cite{Chaloupka2013} has been applied to understand the $T$-dependent local moment\cite{Gretarsson2013} and associated anomalous $T$ dependence of $\chi$ (Ref.~\onlinecite{Johnston2010}) in the PM states of the iron arsenides.

The ICF model is derived from the $\chi$ of an isolated two-level local-moment magnetic system.  Here we apply it to local-moment $d$-electron systems with quenched orbital angular momentum ($L_z=0$).  Spin-orbit coupling generally causes a deviation of $g$ from the free-electron value $g=2$ and the introduction of a paramagnetic Van Vleck orbital susceptibility.  The ground state has energy $E_0\equiv 0$, spin $S_0$ and Zeeman degeneracy $2S_0+1$ and the excited state has energy $E_1$, spin $S_1$ and degeneracy $2S_1+1$.  The spin susceptibility of this two-level system is then
\bse
\label{Eqs:2levelsystem}
\be
\chi_{\rm two-level}(T) = P_0\chi_0(T) + P_1\chi_1(T),
\ee
where the probabilities of occupying state~0 or~1 are given by the Boltzmann distribution as
\be
P_i = \frac{(2S_i+1)e^{-E_i/k_{\rm B}T}}{(2S_0+1) + (2S_1+1)e^{-E_1/k_{\rm B}T}}\quad(i=0,1).
\ee
The susceptibilities of the two levels are given by Curie laws as
\be
\chi_i(T) = \frac{C_i}{T}\quad(i=0,1)
\ee
with Curie constants
\be
C_i = \frac{N_{\rm A}g_i^2S_i(S_i+1)\mu_{\rm B}^2}{3k_{\rm B}}\quad(i=0,1).
\label{Eq:Ci}
\ee
\ese
The phenomenological ICF susceptibility $\chi_{\rm ICF}$ is derived from the above results for an isolated two-level system by making the replacement
\bse
\label{Eqs:models12}
\be
T\to T^\star
\ee
in Eqs.~(\ref{Eqs:2levelsystem}), where\cite{Franz1980}
\be
T^\star =  \sqrt{T^2+T_{\rm sf}^2}\quad ({\rm model~1})
\ee
or\cite{Sales1975}
\be
T^\star =  T+T_{\rm sf}\hspace{0.34in} ({\rm model~2})
\ee
\ese
and $T_{\rm sf}$ is an effective spin fluctuation temperature that reflects the frequency of fluctuations in the spin state of the magnetic ion between $S_0$ and $S_1$.

\begin{table*}
\caption{Parameters obtained from fitting the powder-averaged susceptibility $\chi_{\rm P}(T)$ of $\rm{SrCo_2As_2}$ using the phenomenological interconfigurational flutuation (ICF) model. The fitted parameters are the energy of the  excited triplet states $E_{\rm 1}$, spin fluctuation temperature $T_{\rm sf}$, temperature independent contribution to the magnetic susceptibility $\chi_{0}$ and spectroscopic splitting factor $g_1$ for the excited-state Zeeman levels.  Parameters derived from these are the Curie constant $C_1$ for the excited triplet levels, the Curie constant $C_{\rm ICF}$ and Weiss temperature $\theta_{\rm ICF}$ for the Curie-Weiss-like susceptibility in Eq.~(\ref{Eq:chiICFCW}) followed at high $T$ in the ICF model.} 
\label{Table:ICF}
\begin{ruledtabular}
\begin{tabular}{l c c c c c c c}
	Model     & $E_1/k_{\rm B}$ & $T_{\rm sf}$ & $\chi_0$ & $g_1$ & $C_1$ & $C_{\rm ICF}$  & $\theta_{\rm ICF}$ \\
	     & (K) & (K) & ${\rm (10^{-3}\,cm^3/mol) }$ & & (${\rm cm^3\,K/mol}$) & (${\rm cm^3\,K/mol}$)   &(K)\\
	\hline
	Model~1   & 224(1) & 83.7(7)  & $-0.85(1)$  & 2.25(5) & 1.26 & 0.95  &  $-56$\\
	Model~2   & 468(5) & 183(3)   & $-4.29(6)$  & 4.8(2)  &  5.76 & 4.32  &  $-300$\\
\end{tabular}
\end{ruledtabular}
\end{table*}

\begin{figure}
\includegraphics[width=3.3in]{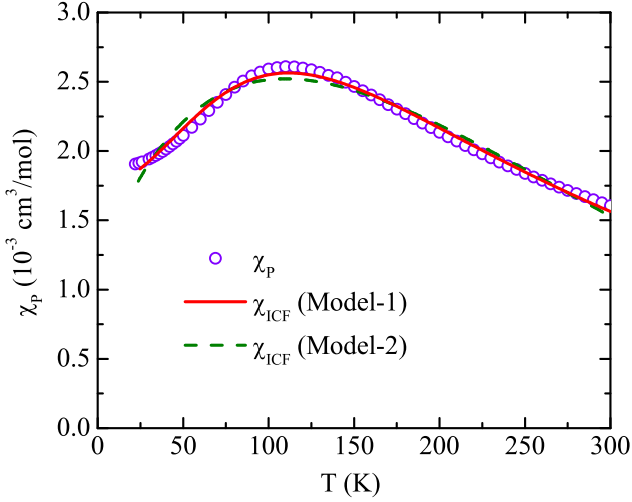}
\caption{(Color online) Powder averaged susceptibility $\chi_{\rm P}$ versus $T$ of $\rm{SrCo_2As_2}$ along with the fit using the two different interconfigurational fluctuation  models in Eqs.~(\ref{Eqs:models12}).}
\label{fig:Figure_IVF}
\end{figure}

We fitted our experimental $\chi(T)$ by
\begin{equation}
\chi(T) = \chi_{0} + \chi_{\rm ICF}(T) ,
\label{Eq:ICF}
\end{equation} 
where $\chi_{0}$ is the $T$-independent orbital susceptibility and $\chi_{\rm ICF}$ is the ICF spin susceptibility.  We tried the combinations $(S_0,S_1) =  (0, \frac{1}{2}),\ (0, 1),\ (0, \frac{3}{2}),\ (\frac{1}{2}, 1)$ and $(1, \frac{3}{2})$, but only the combination $(S_0,S_1) = (0, 1)$ gave a satisfactory fit to the experimental data, for which Eqs.~(\ref{Eqs:2levelsystem}) and~(\ref{Eqs:models12}) yield
\bse
\be
P_1(T) = \frac{3}{3 + e^{E_1/k_{\rm B}T^{\star}}}\quad (S_0=0,\ S_1=1)
\label{Eq:P1T}
\ee
and, using $\chi_0(T) = 0$ because $S_0=0$,
\be
\chi_{\rm ICF}(T) = \frac{3C_1}{(3 + e^{E_1/k_{\rm B}T^{\star}})T^\star}\quad (S_0=0,\ S_1=1).
\label{Eq:chiICF}
\ee

Using the expression for $T^\star$ for model~1 in Eqs.~(\ref{Eqs:models12}), a high-$T$ series expansion of Eq.~(\ref{Eq:chiICF}) gives
\bea
\chi_{\rm ICF}(T) &=& \frac{C_{\rm ICF}}{T}\bigg[1 -\frac{E_1/k_{\rm B}}{4T}\label{Eq:ChiICFS01}\\*
&&\hspace{0.5in}-\ \frac{(E_1/k_{\rm B})^2+8T_{\rm sf}^2}{16 T^2} + {\cal O}\left(\frac{1}{T^3}\right) \bigg],\nonumber
\eea
where
\be
C_{\rm ICF} = \frac{3C_1}{4}
\label{Eq:CICF}
\ee
and $T_{\rm sf}$ first appears in the third term.  The factor of 3/4 arises because in the limit of high $T$ the probability of occupying the triplet excited state is 3/4.  In this limit, Eq.~(\ref{Eq:ChiICFS01}) yields a Curie law
\be
\chi_{\rm ICF}(T) = \frac{C_{\rm ICF}}{T},
\ee
where $C_{\rm ICF}$ is the Curie constant of the two-level system at high~$T$\@.  Expanding the first two terms in Eq.~(\ref{Eq:ChiICFS01}) in a Taylor series according to $1+x\approx 1/(1-x)$ gives a Curie-Weiss-like behavior
\be
\chi_{\rm ICF}(T) = \frac{C_{\rm ICF}}{T-\theta_{\rm ICF}}\qquad(T\gg \theta_{\rm ICF}),
\label{Eq:chiICFCW}
\ee
\ese
where the Weiss temperature is
\bse
\label{Eqs:QICF}
\be
\theta_{\rm ICF} = -\frac{E_1}{4k_{\rm B}}\qquad({\rm model~1})
\label{Eq:QICFMod1}
\ee
with a negative sign (AFM-like).  It is interesting that this expression for $\theta_{\rm ICF}$ in model~1 does not contain the ICF parameter $T_{\rm sf}$.  Carrying out the same steps for model~2 gives a Curie-Weiss law~(\ref{Eq:chiICFCW}) with the same Curie constant but a different Weiss temperature
\be
\theta_{\rm ICF} = -\left(\frac{E_1}{4k_{\rm B}}+ T_{\rm sf}\right)\qquad({\rm model~2}),
\label{Eq:QICFMod1}
\ee
\ese
which for this model does contain $T_{\rm sf}$.

The values of the fitted parameters $\chi_0$, $E_1$, $T_{\rm sf}$ and $g_1$ for models~1 and~2 using $(S_0,S_1)=(0,1)$ are listed in Table~\ref{Table:ICF} and the fits by the two models are shown in Fig.~\ref{fig:Figure_IVF}.  Also listed in Table~\ref{Table:ICF} are the values of $C_1$, $C_{\rm ICF}$, and $\theta_{\rm ICF}$ derived from these parameters using Eqs.~(\ref{Eq:Ci}), (\ref{Eq:CICF}) and~(\ref{Eqs:QICF}), respectively.  Although good fits are obtained using both models, the $\chi_0$ value for the fit by model~2 is unphysically large and negative and even the value for model~1 is unusually negative.  Furthermore, the values of $C_{\rm ICF}$ and $\theta_{\rm ICF}$ are much closer to the Curie-Weiss fit values in Table~\ref{Table:CWFitPars} for model~1 compared to model~2.  However, we only expect the fitting parameters to be semi-quantitative because the ICF model is phenomenological.

\begin{figure}[t]
\includegraphics[width=3.3in]{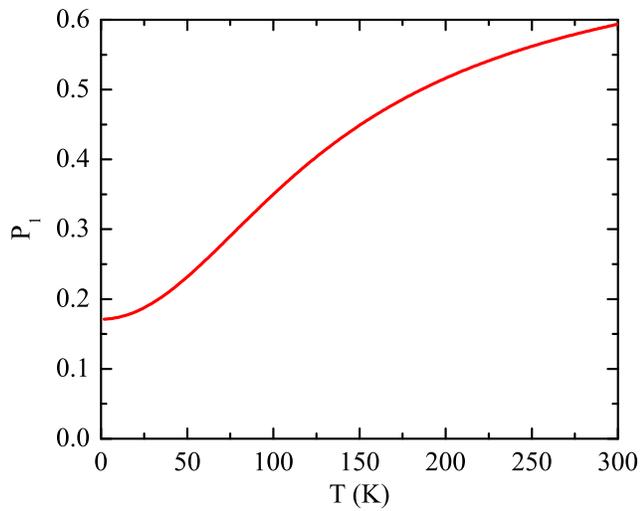}
\caption{(Color online) Temperature $T$ dependence of the fractional population $P_1$ of the excited magnetic state with spin $S_1 = 1$ for model~1 in Eqs.~(\ref{Eqs:2levelsystem}), (\ref{Eqs:models12}) and~(\ref{Eq:P1T}).}
\label{fig:Figure_ICF}
\end{figure}

Figure~\ref{fig:Figure_ICF} shows the $T$ dependence of the fractional population $P_1$ of the magnetic excited triplet state with $S_1=1$ for model~1 and the parameters in Table~\ref{Table:ICF}. Our thermal expansion measurements in Fig.~\ref{Tdep}(a) show that $\rm{SrCo_2As_2}$ exhibits anomalous negative thermal expansion along the $c$-direction and positive thermal expansion in the basal plane, which results in a $a/c$ ratio that is $T$ dependent and increases with increasing $T$ with a dependence similar to that in Fig.~\ref{fig:Figure_ICF} [see Fig.~\ref{Tdep}(c)].  Thus the $T$ dependence of $a/c$ and the population $P_1(T)$ are correlated within this model.

\subsection{$J_1$-$J_2$ Heisenberg Model with $S=1/2$}

\subsubsection{Introduction}

It is well known that low-dimensional Heisenberg antiferromagnets exhibit a $\chi(T)$ that has a broad maximum due to the increasing dynamic short-range AFM correlations on cooling below the temperature of the maximum, in conjunction with the suppression of long-range AFM order due to quantum fluctuations associated with the low spin-lattice dimensionality.\cite{Johnston1997}  In view of the inelastic neutron scattering results that demonstrated the occurrence of dynamic stripe-type AFM correlations in $\rm{SrCo_2As_2}$,\cite{Jayasekara2013} we consider in this section a description of our $\chi(T)$ data in Fig.~\ref{fig:Figure_Mag}, which show a broad maximum, in terms of the $J_1$-$J_2$ Heisenberg model for the two-dimensional (2D) spin-1/2 square lattice.  Due to the Mermin-Wagner theorem,\cite{Mermin1966} a 2D Heisenberg spin lattice cannot exhibit static long-range AFM order at finite~$T$\@. This result is due to fluctuation affects associated with the low dimensionality of the spin lattice.  However, even weak interlayer coupling can induce long-range AFM order, but usually at a temperature significantly below the temperature of the broad peak in $\chi(T)$.\cite{Johnston2011}

The $J_1$-$J_2$ Heisenberg model with spins $S=1/2$ on a square lattice is well studied.  The Hamiltonian in zero field is
\be
{\cal H} = \sum_{\langle ij\rangle}J_1 {\bf S}_i\cdot{\bf S}_j + \sum_{\langle ik\rangle}J_2 {\bf S}_i\cdot{\bf S}_j, 
\ee
where the first sum is over nearest-neighbor spin pairs along the legs of the square lattice and the second sum is over next-nearest-neighbor spin pairs along the diagonals of the squares, where positive $J_1$ and $J_2$ correspond to AFM interactions and negative ones to FM interactions.  A stripe-type AFM phase as observed in the LaFeAsO and $A{\rm Fe_2As_2}$ compounds occurs classically for $J_2>0$ and $J_2/|J_1| > 1/2$ where $J_1$ can be either positive (AFM) or negative (FM).\cite{Johnston2010}  Early in the Fe-based high-$T_{\rm c}$ superconductivity field, this (effective) frustration model was used to explain the stripe-type AFM ordering observed in the 1111-type and 122-type iron arsenides in both the itinerant\cite{Yildirim2008} and local moment\cite{Si2008} descriptions of the magnetism.  The same stripe-type AFM ordering is predicted from band theory without reference to exchange interactions.\cite{Mazin2008,Dong2008}

Our recent inelastic neutron scattering measurements on $\rm{SrCo_2As_2}$ indicate the presence of stripe-type AFM correlations with $J_1<0$ and $J_2/J_1\approx -1$,\cite{Jayasekara2013} which is consistent with the above classical constraints for stripe-type AFM correlations, and further indicated that the correlations are quasi-two-dimensional.  We therefore proceed to analyze our $\chi(T)$ data for $\rm{SrCo_2As_2}$ in terms of the frustrated 2D $J_1$-$J_2$ spin-1/2 Heisenberg square lattice model.

\subsubsection{Determining $J_1$ and $J_2$ from $\chi(T)$ Data}

In order to estimate the values of the exchange constants $J_1$ and $J_2$, one can fit $\chi(T)$ by the high-$T$ series expansion (HTSE) for the spin susceptibility of this model which is given for $J_1<0$ and $J_2>0$ by\cite{Rosner2003}
\be
\frac{\chi|J_1|}{Ng^2\mu_{\rm B}^2} = \frac{|J_1|}{k_{\rm B}T}\sum_{n=0}^9\left(\frac{-|J_1|}{k_{\rm B}T}\right)^n \sum_{m=0}^n c_{m,n}\left(\frac{J_2}{J_1}\right)^m,
\label{Eq:HTSE}
\ee
where $N$ is the number of spins and a table of the $c_{m,n}$ values is given in Ref.~\onlinecite{Rosner2003}.  For a given value of $J_2/J_1$, we converted the sum over~$n$ into a Pad\'e approximant ${\cal P}[p,q]$ with $[p,q]=[4,4]$, which is a ratio of two fourth-order polynomials in $|J_1|/k_{\rm B}T$ ($p=q=4$), where the coefficients of the two polynomials are solved for exactly in terms of the original HTSE coefficients, which in turn results in an expression $\chi_{\rm Pade}(T)$ for the susceptibility.  The Pad\'e approximant reformulation of the HTSE prediction for $\chi$ decreases the minimum temperature at which the HTSE can be used to fit experimental data.  

\begin{figure}[t]
\includegraphics[width=3.3in]{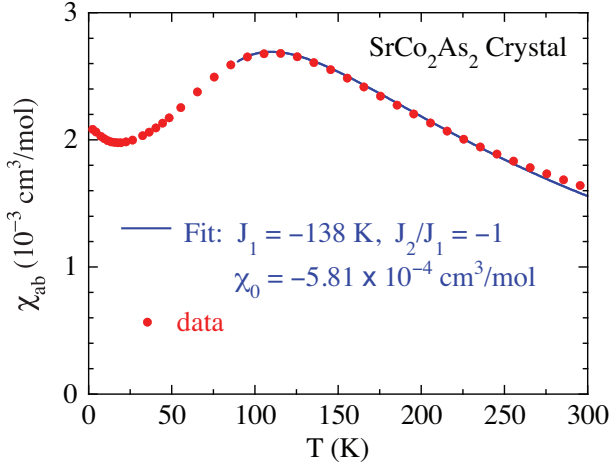}
\caption{(Color online) A fit of the $ab$-plane susceptibility $\chi_{ab}(T)$ data for $\rm{SrCo_2As_2}$ from Fig.~\ref{fig:Figure_Mag} (filled red circles) by the $J_1$-$J_2$ spin-1/2 Heisenberg model on a square lattice according to Eq.~(\ref{Eq:HTEChiFit}) (blue curve).  The fitting parameters are given in the figure where $g=2$ was assumed.}
\label{Fig:SrCoAs_ChiFitPade44PRB}
\end{figure}

We fitted our $\chi(T)$ data by
\be
\chi(T) = \chi_0 + \chi_{\rm Pade}(T,|J_1|,J_2/J_1).
\label{Eq:HTEChiFit}
\ee
A fit of the experimental $\chi_{ab}(T)$ data in Fig.~\ref{fig:Figure_Mag} by Eq.~(\ref{Eq:HTEChiFit}) for $J_1<0$ and $J_2/J_1=-1$ was obtained, as dictated by the inelastic neutron scattering measurements,\cite{Jayasekara2013} and we also fixed $g=2$.  The fit is shown in Fig.~\ref{Fig:SrCoAs_ChiFitPade44PRB}, where the fitting parameters $\chi_0$ and $J_1$ are given in the figure.  The high-temperature series and Pad\'e approximant derived from it are not expected to be valid to temperatures much below the temperature of the broad maximum, so we cut off the fit as shown.  The value of $\chi_0$ is perhaps too negative to be physical, which suggests that the model may not include all relevant exchange couplings.

\subsubsection{Extended 3D $J_1$-$J_2$-$J_c$ Spin $S=1/2$ Heisenberg Model}

The above fit of $\chi(T)$ by the $J_1$-$J_2$ model ignores the magnetic coupling between adjacent stacked Co square lattice layers along the $c$-axis.  To include this coupling, we utilize the above intralayer $\chi_{\rm Pade}(T)$ in Eq.~(\ref{Eq:HTEChiFit}) as determined from Eq.~(\ref{Eq:HTSE}), but augment the $J_1$-$J_2$ Heisenberg square lattice model by adding a nearest-neighbor $c$-axis Heisenberg coupling between the stacked square lattices using molecular field theory (MFT) as follows, where we utilize the fact that all Co spins are identical and crystallographically equivalent in ${\rm SrCo_2As_2}$.

The time- and space-average magnetic moment component ${\mu}_i$ in the direction of the applied field ${\bf H}$ is
\bse
\be
\mu_i = \frac{\chi_{\rm Pade}}{N}(H + H_{{\rm exch}\,c\,i}),
\label{Eq:muiMFT}
\ee
where $H_{{\rm exch}\,c\,i}$ is the exchange field seen by $\vec{\mu}_i$ due to the two moments in the two adjacent layers along the $c$-axis.  In MFT, one writes 
\be
H_{{\rm exch}\,c\,i} = \lambda_c \mu_i,
\label{Eq:Hexchlambda}
\ee
where $\lambda_c$ is the MFT coupling constant of $\vec{\mu}_i$ to the two moments in adjacent layers along the $c$-axis.  Substituting Eq.~(\ref{Eq:Hexchlambda}) into~(\ref{Eq:muiMFT}) and solving for $\mu_i$ gives
\be
\mu_i = \frac{\frac{\chi_{\rm Pade}}{N}H}{1-\frac{\chi_{\rm Pade}}{N}\lambda_c},
\ee
so the spin susceptibility per spin is
\be
\chi_i^{\rm spin} = \frac{\mu_i}{H} = \frac{\chi_{\rm Pade}/N}{1-\frac{\chi_{\rm Pade}}{N}\lambda_c},
\label{Eq:chilambda}
\ee
and for $N$ spins one then obtains
\be
\chi^{\rm spin} = \frac{\chi_{\rm Pade}}{1-\frac{\chi_{\rm Pade}}{N}\lambda_c}.
\label{Eq:chilambda2}
\ee
Taking the reciprocal of each side of this equation gives
\be
\frac{1}{\chi^{\rm spin}} = \frac{1}{\chi_{\rm Pade}} - \frac{\lambda_c}{N}.
\ee
\ese
Thus the MFT coupling parameter $\lambda_c$ just shifts the entire inverse susceptibility versus~$T$ curve vertically by an amount $-\lambda_c/N$, and therefore does not change the temperature of the maximum in $\chi(T)$ at about 115~K in Fig.~\ref{Fig:SrCoAs_ChiFitPade44PRB}.

Now we express $\lambda_c$ in terms of an exchange constant $J_c$ between a spin and a nearest-neighbor spin in an adjacent layer.  The part of the Heisenberg Hamiltonian associated with the interaction of spin {\bf S}$_i$ with the two nearest-neighbor spins {\bf S}$_j$ in adjacent layers is
\be
{\cal H}_i = \frac{1}{2}J_c{\bf S}_i\cdot \sum_{j=1}^2 {\bf S}_j = \frac{J_c}{2g^2\mu_{\rm B}^2}\vec{\mu}_i\cdot \sum_{j=1}^2 \vec{\mu}_j,
\label{Eq:CalHi}
\ee
where the factor of 1/2 arises because the exchange energy is equally shared by each spin within a pair of interacting spins.  The magnetic moment $\vec{\mu}$ and spin {\bf S} vectors are related by $\vec{\mu} = -g{\bf S}\mu_{\rm B}$, where the minus sign arises from the negative sign of the charge on the electron.  In MFT, one writes the time- and space-average of the energy per spin $E_i$ in the PM state as
\be
E_i = \frac{J_c}{g^2\mu_{\rm B}^2}\mu_i\mu_j \equiv -\frac{1}{2}\mu_i H_{{\rm exch}\,c\,i},
\label{Eq:Ei}
\ee
where the factor of 1/2 on the right-hand side arises because when calculating the exchange field, all of the interaction energy $E_i$ between moments $\vec{\mu}_i$ and $\vec{\mu}_j$ is assigned to the neighbor $\vec{\mu}_j$ of $\vec{\mu}_i$.  Then, recognizing that the average moments satisfy $\mu_i=\mu_j$ in the PM state, the exchange field associated with the $c$-axis interactions is found from Eq.~~(\ref{Eq:Ei}) to be
\be
H_{{\rm exch}\,c\,i} = -\frac{2J_c}{g^2\mu_{\rm B}^2}\mu_i.
\label{Eq:HexchJc}
\ee
Comparing Eqs.~(\ref{Eq:HexchJc}) and~(\ref{Eq:Hexchlambda}) gives $\lambda_c$ in terms of $J_c$ as
\be
\lambda_c = -\frac{2J_c}{g^2\mu_{\rm B}^2}.
\label{Eq:lambdacJc}
\ee
Substituting Eq.~(\ref{Eq:lambdacJc}) into~(\ref{Eq:chilambda2}) gives 
\be
\chi^{\rm spin}(T) = \frac{\chi_{\rm Pade}(T)}{1+ \frac{2J_c\chi_{\rm Pade}(T)}{Ng^2\mu_{\rm B}^2}}.
\label{Eq:chiJc}
\ee

From Eq.~(\ref{Eq:chiJc}), an AFM (positive) $J_c$ suppresses the susceptibility, whereas a FM (negative) $J_c$ enhances it, as expected.  Our inelastic neutron scattering results indicated an AFM Co-Co interlayer interaction in $\rm{SrCo_2As_2}$.\cite{Jayasekara2013}  Therefore the anomalously negative value of $\chi_0$ derived in the previous section that suppresses $\chi$ probably arises at least in part from the presence of the $c$-axis AFM Co-Co interactions that are neglected in the $J_1$-$J_2$ model.  Using Eq.~(\ref{Eq:chiJc}), in order to produce the same effect on $\chi(T)$ at its maximum as $\chi_0=-5.8\times10^{-4}\,{\rm cm^3/mol}$ (Fig.~\ref{Fig:SrCoAs_ChiFitPade44PRB}) would require $J_c/k_{\rm B} = 460$~K, which according to our inelastic neutron scattering data (Ref.~\onlinecite{Jayasekara2013}) is much too large compared to $|J_1|,\,J_2 = 138$~K and should therefore be considered an upper limit.  We conclude that within the present model, a significant diamagnetic $\chi_0$ value as well as a significant AFM $c$-axis coupling are present.

Finally, we add a caveat that our analyses of the magnetic properties of AFM $\rm{CaCo_2As_2}$ indicate that the magnetism is itinerant,\cite{Anand2013} whereas the above analyses of $\chi(T)$ for $\rm{SrCo_2As_2}$ were carried out within a local-moment picture.

\section{\label{Sec:NMR} NMR Spectroscopy}

\begin{figure}[tb]
\includegraphics[width=3.3in]{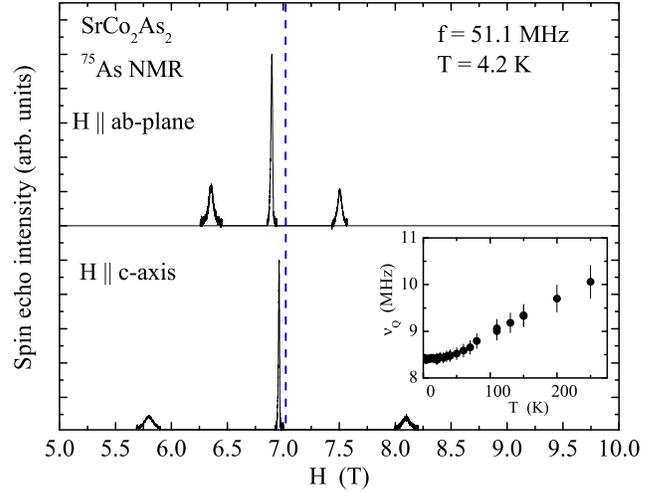} 
\caption{(Color online) Field-swept $^{75}$As-NMR spectra of $\rm{SrCo_2As_2}$ at temperature $T = 4.2$~K for magnetic fields $H$ perpendicular and parallel to the $c$-axis. The vertical dashed blue line corresponds to the zero-shift ($K = 0$) position.  Inset: $^{75}$As nuclear quadruple frequency $\nu_{\rm{Q}}$ versus $T$\@.}
\label{fig:As-spectrum}
\end{figure}

Figure~\ref{fig:As-spectrum} shows field-swept $^{75}$As-NMR spectra of SrCo$_2$As$_2$ at $T$ = 4.2 K for two magnetic field directions of $H$ $\parallel$ $c$-axis and $H$ $\parallel$ $ab$-plane.  The spectra exhibit features typical of a nuclear spin $I = 3/2$ with Zeeman and quadrupolar interactions, which result in a sharp central transition and two satellite peaks split by the quadrupolar interaction of the $^{75}$As nucleus with the local electric field gradient (EFG).  The principle axis of the EFG at the As site in tetragonal SrCo$_2$As$_2$ is found to be along the crystal $c$~axis as for members of the $A{\rm Fe_2As_2}$ family that have tetragonal symmetry,\cite{Kitagawa2009, Kitagawa2008, Baek2008, Ning2009, Urbano2010} since in our CoAs-based compound and in the tetragonal FeAs-based materials the As site has a local fourfold rotational symmetry about the $c$~axis.  The spectra for SrCo$_2$As$_2$ for both $H$ directions did not show any obvious magnetic broadenings at any temperature between 1.6 and 275~K, which indicates that static magnetic ordering does not occur in this compound above 1.6~K\@.

The inset of Fig.~\ref{fig:As-spectrum} shows the temperature dependence of the nuclear quadrupole frequency $\nu_{\rm{Q}} =  eQV_{ZZ}/2h$ where $Q$ is the electric quadrupole moment of the $^{75}$As nucleus, $V_{ZZ}$ is the principal-axis EFG at the As site and $h$ is Planck's constant.  $V_{ZZ}$ arises from  hybridization between the As-4$p$ and Co-3$d$ orbitals with an additional contribution from the  noncubic part of the spatial distribution of surrounding ions.  The  $\nu_{\rm{Q}}$  decreases by about 15\% on cooling, from 10~MHz at 250~K to 8.45~MHz at 1.6 K\@.  A decrease of $\nu_{\rm{Q}}$  by $\sim15$\% on lowering $T$ from 280 to 140~K was also observed in BaFe$_2$As$_2$ (Ref.~\onlinecite{Kitagawa2009}) where $\nu_{\rm{Q}}$ is much smaller in BaFe$_2$As$_2$ than in SrCo$_2$As$_2$.  These behaviors sharply contrast with that observed for SrFe$_2$As$_2$ where $\nu_{\rm{Q}}$ \emph{increases} from 2.0~MHz at 300~K to 2.6~MHz at $\sim 210$~K\@.\cite{Kitagawa2009}

\begin{figure}[tb]
\includegraphics[width=3.in]{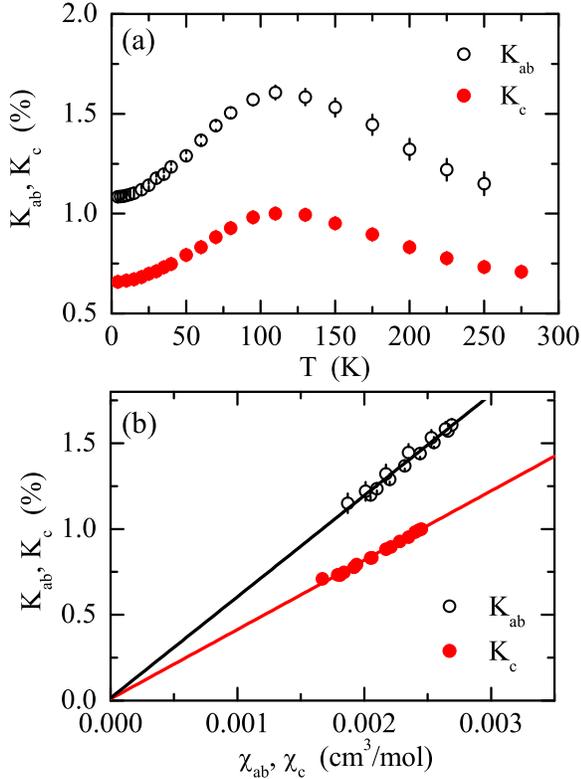} 
\caption{(Color online) (a) Temperature $T$-dependence of the $^{75}$As NMR shifts $K_{ab}$ and $K_{c}$ for $\rm{SrCo_2As_2}$. (b) $K$ versus $\chi$ plots for the corresponding $ab$ and $c$ components of $K$ and $\chi$ with $T$ as an implicit parameter for $35~{\rm K} \leq T \leq 275$~K\@.  The linear fits are not forced to go through the origin; however, the fitted $y$-intercepts are almost zero, as shown.}
\label{fig:K-chi}
\end{figure}

Figure~\ref{fig:K-chi}(a) shows the $T$ dependences of the NMR shifts $K_{ab}$ and $K_{c}$ for $H$ parallel to the $ab$ plane and parallel to the $c$~axis, respectively, where the second-order quadrupole shift was corrected for in $K_{ab}(T)$.  As $T$ decreases below room temperature, both $K_{ab}$ and $K_{c}$ increase and show broad maxima at the about the same temperature of $\approx 115$~K, similar to the $\chi(T)$ behavior in Fig.~\ref{fig:Figure_Mag}, and then decrease monotonically at lower~$T$ without showing upturns.  The small upturns in $\chi(T)$ observed at low $T$ in Fig.~\ref{fig:Figure_Mag} are therefore not intrinsic and evidently arise from a small amount of a PM impurity that is not saturable in fields up to 5.5~T (see Fig.~\ref{fig:Figure_MH}).

Figure~\ref{fig:K-chi}(b) shows $K_{ab}$ and $K_c$ from Fig.~\ref{fig:K-chi}(a) versus $\chi_{ab}$ and $\chi_c$ from Fig.~\ref{fig:Figure_Mag}, respectively, with $T$ as an implicit parameter, for $T > 35$~K to avoid the extrinsic low-temperature upturns in $\chi(T)$.   Both $K_{ab}$ and $K_{c}$ vary linearly with the corresponding $\chi$ and, from the respective slopes, the hyperfine coupling constants are estimated to be $A_{ab} = (65.9\pm2.8)$~kOe/$\mu_{\rm B}$  and $A_{c} = (45.0\pm1.0)$~kOe/$\mu_{\rm B}$, respectively.  These values are much larger than $A_{ab} = 29.3$~kOe/$\mu_{\rm B}$  and $A_{c}$ = 20.4 kOe/$\mu_{\rm B}$ for SrFe$_2$As$_2$, respectively.\cite{Kitagawa2009}  Since the hyperfine coupling is mainly due to hybridization between the $4s$ and $4p$ orbitals of the As atoms with the $3d$ orbitals of the Co atoms, the large hyperfine couplings indicate a strong hybridization between these orbitals.  This is consistent with the large $\nu_{\rm{Q}}$ for $^{75}$As in SrCo$_2$As$_2$ because the $\nu_{\rm{Q}}$ is also affected by the strength of the hybridization between the As-$4p$ and Co-$3d$ orbitals.  The $y$-intercepts of the respective linear fits of $K$ versus $\chi$ in Fig.~\ref{fig:K-chi}(b) are nearly zero.  This  indicates that the NMR shifts originate mainly from the spin susceptibility with almost negligible net contribution from the orbital (chemical) paramagnetic Van Vleck and diamagnetic conduction electron Landau susceptibilities.  Therefore, we identify the measured NMR shifts as Knight shifts.

\begin{figure}[tb]
\includegraphics[width=3.4in]{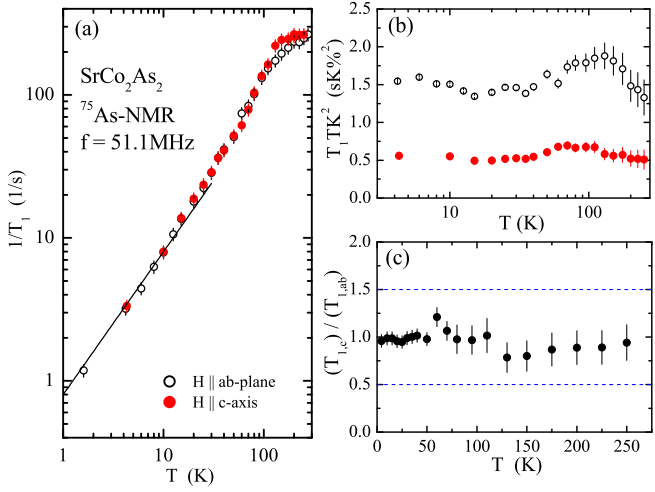}
\caption{(Color online) (a) Temperature dependence of 1/$T_1$ for $\rm{SrCo_2As_2}$ for both magnetic field directions, $H \parallel c$-axis and $H \parallel ab$-plane.  The straight line shows the metallic Korringa dependence $1/T_1 \propto T$\@.  (b) $T$-dependence of $T_{\rm 1}TK^{\rm 2}$. The symbol designations are the same as in panel~(a).  (c)~\mbox{$T$~dependence} of the ratio $r\equiv T_{1,c}/T_{1, ab}$. }
\label{fig:T1}
\end{figure}

The $^{75}$As nuclear spin-lattice relaxation rate 1/$T_{\rm 1}$ at each $T$ is determined by fitting the nuclear magnetization $M$ versus time $t$ dependence of the central line after saturation using the double-exponential function $1-M(t)/M({\infty}) = 0.1\exp(-t/T_1) + 0.9\exp(-6t/T_1$) as expected for the central line of the spectrum of the $^{75}$As ($I = 3/2$) nucleus, where $M(t)$ and $M({\infty})$  are the nuclear magnetization at time $t$ after saturation and the equilibrium nuclear magnetization at $t \to \infty$, respectively.  Figure~\ref{fig:T1}(a) shows $1/T_{\rm 1}$ versus~$T$ for magnetic fields parallel and perpendicular to the $c$~axis.  Both 1/$T_1$ measurements show very similar $T$~dependences which consist of a weak $T$-dependence at high temperatures above $\sim 150$~K followed by a monotonic decrease at lower~$T$\@.

Within a Fermi liquid picture, 1/$T_1T$  is proportional to the square of the density of states at the Fermi energy ${\cal D}(E_{\rm F}$).  On the other hand, the Knight shift is proportional to the Pauli spin susceptibility $\chi^{\rm Pauli} = \mu_{\rm B}^2{\cal D}(E_{\rm F}$).  By looking at the behavior of $T_{\rm 1}TK^2$ versus $T$, one can therefore test the  applicability of the Fermi liquid theory to our system which predicts the Korringa relation $T_{\rm 1}TK^2$ = const.  As shown in Fig.~\ref{fig:T1}(b), $T_{\rm 1}TK^2$  for both $H$ directions are indeed nearly independent of~$T$ for $\rm{SrCo_2As_2}$.  This means that the conduction carriers form a Fermi liquid.  The $T^{\rm 2}$ dependence of $\rho(T)$ observed below 38~K in the inset of Fig.~\ref{fig:Figure_Res} above is another indication of Fermi liquid behavior. 

Next, we discuss the relationship between the anisotropy of our $1/T_{\rm 1}$ data and the anisotropy of the Co spin fluctuations. Here we follow the same procedure previously performed on Fe pnictides.\cite{Kitagawa2009, Kitagawa2010, Hirano2012}  In general, $1/T_1$ can be described in terms of the fluctuating hyperfine field perpendicular to the applied magnetic field at the NMR angular frequency $\omega_{\rm N}$ according to 
\begin{eqnarray}
 \left (\frac{1}{T_1}\right)_z 
& = & \frac{\gamma_{\rm N}^2}{2}  \int_{-\infty}^{\infty} \Big[\langle H_{{\rm hf},x}(t),H_{{\rm hf},x}(0)\rangle \nonumber \\*
 && +\ \langle H_{{\rm hf},y}(t),H_{{\rm hf},y}(0)\rangle\Big] \exp(i \omega_{\rm N}t)\,dt \label{eqn:T1} \\*
& = &  \gamma_{\rm N}^{2}\sum_{\bf q}\Big[\big|H_{{\rm hf},x}({\bf q}, \omega_{\rm N})\big|^2 +  \big|H_{{\rm hf},y}({\bf q}, \omega_{\rm N})\big|^2 \Big] ,  \nonumber
\end{eqnarray}
where the $z$ direction corresponds to the external field direction and $| X({\bf q}, \omega_{\rm N}) | ^2$ denotes the {\bf q}-dependent power spectral density at $\omega_{\rm N}$ of a time-dependent random variable $X(t)$.

Defining the hyperfine coupling tensor as $\tilde {A}$, the hyperfine fields at the As site can be described as the sum of contributions from the four nearest-neighbor Co spins as\cite{Kitagawa2008, Hirano2012}
\begin{equation}
{\bf H}_{\rm hf}({\bf q}, \omega_{\rm N}) = \sum_{i=1}^4 {\tilde{A}_i} {\bf S}_i({\bf q}, \omega_{\rm N}) = \tilde{A} {\bf S}_i({\bf q}, \omega_{\rm N}),
\end{equation}
where
\begin{eqnarray}
\tilde{A}  = \left (
\begin {array}{ccc}
A_a 	~ D  ~	 B_1\\
D ~	A_b ~      B_2\\
B_1 ~      B_2 ~       A_c
\end {array}
\right).
\end{eqnarray}
Here $\tilde {A}$ is described in orthorhombic axis (stacked Co square-lattice) notation for consistency with the previous discussion on Fe pnictides.\cite{Kitagawa2008}  We consider the cases where $A_\alpha$ ($\alpha = a,\ b,\ c$), $B_i$  ($i$ = 1,2) and $D$ in $\tilde {A}$ are, respectively, the diagonal components along the $\alpha$ direction with {\bf q} = (0,0), the off-diagonal components related to stripe AFM correlations with ${\bf q} = (\pi, 0)$ or $(0, \pi$), and the off-diagonal components associated with N\'eel-type spin correlations with ${\bf q} = (\pi, \pi)$. \cite{Kitagawa2010}  Defining the amplitude of the spin fluctuation spectral density as ${\cal S}_\alpha\equiv \sqrt{|S_\alpha({\bf q},\omega_{\rm N})|^2}\ (\alpha = a,\ b,\ c)$, using the above equations and writing  $A_{ab} \equiv  A_{a} = A_{b}$ and ${\cal S}_{ab} \equiv {\cal S}_a = {\cal S}_b$ appropriate to the tetragonal symmetry of ${\rm SrCo_2As_2}$, the ratio
\be
r\equiv \frac{T_{1,c}}{T_{1, ab}}
\ee
is obtained as\cite{Hirano2012}
\begin{subnumcases}
{\textstyle r =}
0.5+0.5\left( \frac{A_{c}{\cal S}_{c}}{A_{ab}{\cal S}_{ab}}\right) ^{2}   \quad{\rm for}\ {\bf q} = (0,0), & \label{T1RA} \\
0.5+ \left( \frac{{\cal S}_{ab}}{{\cal S}_{c}}\right) ^{2} \quad{\rm for}\  {\bf q} = (\pi,0)\; {\rm or}\; (0,\pi), & \label{T1RB}  \\
0.5 \hspace{1.24in} {\rm for}\ {\bf q} = (\pi,\pi). & \label{T1RC}
\end{subnumcases}

As plotted in Fig.~\ref{fig:T1}(c), $r$ is almost constant versus $T$ with a value $r\approx 1$.  From our inelastic neutron scattering measurements, the dominant AFM fluctuations/correlations are stripe-type with ${\bf q} = (\pi,0)$ or~$(0,\pi)$ in Co square-lattice notation.\cite{Jayasekara2013}  According to Eq.~(\ref{T1RB}), the value $r\approx 1$ observed in Fig.~\ref{fig:T1}(c) then arises from anisotropic spin fluctuations with ${\cal S}_{ab}\approx 0.71\,{\cal S}_c$.  This means that stripe-type AFM Co spin fluctuations along the $c$-axis are stronger than in the $ab$ plane.  An anisotropy in spin fluctuations is also observed in various Fe-based superconductors in the paramagnetic state near $T_{\rm N}$, but ${\cal S}_{ab}$ is always larger than ${\cal S}_c$,\cite{Kitagawa2009, Nakai2012} in contrast to our result for ${\rm SrCo_2As_2}$.  N\'eel-type AFM spin fluctuations with ${\bf q} = (\pi,\pi)$ can be ruled out because according to Eq.~(\ref{T1RC}), that would require $r=0.5$ in conflict with our measurements in Fig.~\ref{fig:T1}(c) that give $r\approx 1$.

Our $1/T_1$ data can also be analyzed with Eq.~(\ref{T1RA}) solely in terms of FM spin fluctuations with ${\bf q} = (0, 0)$.  Using the ratio of the hyperfine fields $A_c/A_{ab}$ = 45.0/65.9 = 0.653 determined from the $K$-$\chi$ plot in Fig.~\ref{fig:K-chi}(b), $r\approx 1$ can be reproduced with anisotropic spin fluctuations with ${\cal S}_{ab} = 0.65\, {\cal S}_c$.  Interestingly, this analysis also gives the same nature of the anisotropy of the spin fluctuations as discussed above for stripe-type AFM fluctuations.  Therefore one cannot uniquely determine the nature of the spin fluctuations utilizing NMR data alone.  The {\bf q}-dependent static susceptibility $\chi({\bf q})$ obtained from electronic structure calculations showed a peak at ${\bf q} = (0,0)$ as well as at ${\bf q} = (\pi,0)$ and $(0,\pi)$,\cite{Jayasekara2013} so spin fluctuations may well occur with peaks at both the FM and stripe-type AFM wave vectors.

From our NMR data on SrCo$_2$As$_2$ crystals we conclude that the conduction carriers form a Fermi liquid that exhibits significant FM and/or AFM spin fluctuations.  The presence of strong AFM spin fluctuations at low temperatures has been specifically demonstrated from our inelastic neutron scattering measurements of single crystals,\cite{Jayasekara2013} and FM correlations from the strongly-enhanced $\chi$ discussed above in Sec.~\ref{Sec:Stoner}.

\section{\label{Sec:MagneticNeuts} Single-Crystal Neutron Diffraction}

\begin{figure}[t]
\includegraphics[width=3.3in]{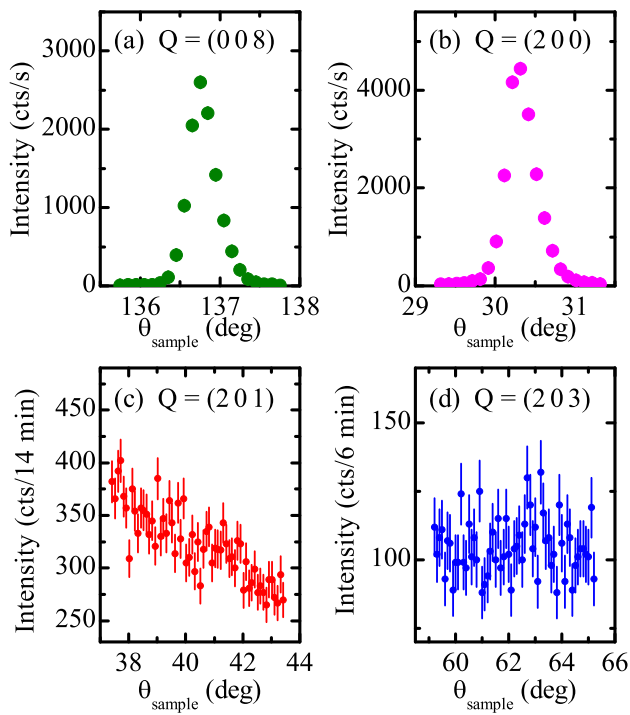}
\caption{(Color online) Single-crystal neutron diffraction measurements at $T = 3$~K searching for A-type AFM order in SrCo$_2$As$_2$. The (0~0~8) and (2~0~0) nuclear Bragg peaks are shown in panels~(a) and~(b), respectively.  The neutron intensity measured as the sample is rocked through the (2~0~1) and (2~0~3) positions at which magnetic Bragg peaks would be observed for A-type AFM ordering is shown in panels~(c) and~(d), respectively.  The calculated values of $\theta_{\rm sample}$ of this potential type of AFM ordering are $40.4^\circ$ in~(c) and $62.2^\circ$ in~(d), respectively, which are near the middle of the respective scans. }
\label{Fig:Neutrons_Mag_Bragg}
\end{figure}

The above thermal expansion, $\rho(T)$, $C_{\rm p}(T)$, $\chi(T)$ and NMR measurements showed no evidence for any phase transitions from~1.3 to~300~K in SrCo$_2$As$_2$.  This result is consistent with the neutron scattering measurements on SrCo$_2$As$_2$ crystals\cite{Jayasekara2013} that showed no evidence for long-range AFM order at the propagation vector of the stripe-type AFM order found\cite{Johnston2010} in the 122-type iron arsenides.  However, the inelastic neutron scattering measurements did show strong dynamic AFM correlations in SrCo$_2$As$_2$ at this propagation vector.\cite{Jayasekara2013}  On the other hand, neutron diffraction measurements on CaCo$_2$As$_2$ confirmed the presence of A-type long-range AFM order characterized by magnetic Bragg peaks at reciprocal lattice points ($h~0~\ell$) with $h + \ell = 2n+1$ [e.g.\ (2~0~1) and (2~0~3)] with an upper limit of 0.6~$\mu_{\rm{B}}$/Co for the ordered moment.\cite{Anand2013}

To complement the above null result for the occurrence of long-range stripe-type AFM ordering in SrCo$_2$As$_2$, here we searched for evidence of A-type\cite{Anand2013} or G-type\cite{SinghY2009} long-range AFM order in a SrCo$_2$As$_2$ crystal using single-crystal neutron diffraction at the base temperature of $T\approx3$~K and did not find any such evidence.  Figure~\ref{Fig:Neutrons_Mag_Bragg} shows scans through both nuclear [Figs.~\ref{Fig:Neutrons_Mag_Bragg}(a) and (b)] and A-type AFM [Figs.~\ref{Fig:Neutrons_Mag_Bragg}(c) and (d)] Bragg peak positions.  There is no evidence of scattering above background at the latter positions.  Note the factor of $\sim 15,000$ finer scale of the ordinates for the potential AFM peak positions in Figs.~\ref{Fig:Neutrons_Mag_Bragg}(c,d) compared to those of the nuclear reflections in Figs.~\ref{Fig:Neutrons_Mag_Bragg}(a,b).  The small increase in scattering with decreasing angle in Fig.~\ref{Fig:Neutrons_Mag_Bragg}(c) is likely due to some degree of preferred orientation in the Al sample~can since the scattering angle is close to that for the Al (2~0~0) powder line.  Measurements to check for G-type long-range AFM order are complicated by the fact that the AFM Bragg reflections are coincident with nuclear Bragg peaks.  However, we found no evidence of enhanced scattering at the (1~0~1) position of the strongest G-type AFM reflection (not shown).

\section{\label{Sec:GndState} Ground State of ${\rm\bf SrCo_2As_2}$}

Our x-ray and neutron diffraction, $\rho(T)$, $M(H,T)$, $\chi(T)$, $C_{\rm p}(T)$ and NMR measurements reveal no evidence for any phase transitions in ${\rm SrCo_2As_2}$ from 300~K down to 1.3~K\@.  On the other hand, the temperature $T_{\rm max}\approx 115$~K at which the broad maximum in $\chi(T)$ occurs, the Weiss temperature $\theta_{\rm p}\sim -150$~K in the Curie-Weiss law fits from 200 to 300~K in Table~\ref{Table:CWFitPars}, the large values of $|J_1|=J_2=138$~K in Fig.~\ref{Fig:SrCoAs_ChiFitPade44PRB} obtained by fitting the $\chi(T)$ data by the $J_1$-$J_2$ model, and the observation of strong stripe-type AFM spin correlations in neutron scattering measurements\cite{Jayasekara2013} demonstrate that AFM spin correlations are present and are strong.  However, long-range AFM order is not observed above 1.3~K\@.  A measure of the ratio of the energy scale of the AFM spin correlations to the long-range AFM transition temperature $T_{\rm N}<1.3$~K is therefore
\be
f = \frac{|\theta_{\rm p}|}{T_{\rm N}} \gtrsim 100.
\label{Eq:frustpar}
\ee

In insulating local-moment systems, $f$ is called the frustration parameter.\cite{Ramirez1994}  In such systems in which frustration prevents long-range AFM ordering, a ratio $f\gtrsim100$ and the absence of static magnetic ordering of any kind occurring in the accessible temperature range point towards a quantum spin-liquid (QSL) ground state, in which quantum fluctuations prevent static long-range magnetic ordering from occurring.  Thus to establish a probable quantum spin-liquid ground state\cite{Balents2010} one needs to demonstrate (1) the presence of spin interactions and have a measure of its strength such as $\theta_{\rm p}$, which establishes the presence of a spin liquid instead of a spin gas, (2) that static magnetic ordering does not occur in the accessible temperature range, (3) that a relation such as Eq.~(\ref{Eq:frustpar}) is satisfied, and (4) that the lattice symmetry is not broken by the putative QSL ground state.  An extensively studied example of an insulator containing Cu$^{+2}$ spins-1/2 that satisfies these conditions is ${\rm ZnCu_3(OH)_6Cl_2}$ (or with deuterium replacing H), which is called herbertsmithite.\cite{Helton2007, Olariu2008, deVries2009, Han2012, Yan2011}

The concept of a QSL ground state has been applied in the past to insulating spin $S=1/2$ local-moment systems.\cite{Balents2010}  However, some itinerant spin systems such as the iron pnictides and chalcogenides and related materials are moderately correlated electron systems and sometimes have both itinerant and local-moment characteristics.\cite{Johnston2010, Dai2012, Dagotto2013, Pandey2012, Lamsal2013}  In such cases we suggest that one may be able to apply the same four criteria above with respect to a possible QSL ground state.  In particular, since we have demonstrated that the above four  conditions are satisfied for metallic ${\rm SrCo_2As_2}$, we suggest that this compound has a quantum spin-liquid ground state.  An alternative and very rare ground state for metallic strongly-correlated frustrated $d$-electron systems is a heavy-fermion ground state as observed in the cubic spinel-structure metallic compound ${\rm LiV_2O_4}$.\cite{Kondo1997,Johnston2000, Lee2001}  This compound shows local-moment Curie-Weiss behavior at high~$T$ and crosses over to the heavy-fermion ground state below about 10~K.

\section{\label{Sec:Summary} Summary and Conclusions}

A comprehensive study of the crystallographic, electronic, magnetic and thermal properties of SrCo$_2$As$_2$ single crystals is reported.  The room-temperature crystal structure was confirmed to be the body-centered tetragonal ${\rm ThCr_2Si_2}$-type ($I4/mmm$) structure with an As--As interlayer distance intermediate between the uncollapsed- and collapsed-tetragonal values in BaCo$_2$As$_2$ and CaCo$_2$As$_2$, respectively.  No phase transitions of any kind were detected between 1.3 and 300~K in any of our measurements, in spite of the strong negative thermal expansion of the $c$~axis from 7.4 to 300~K and the negative volume expansion from~7.4 to $\approx 80$~K\@.  The $a$-axis thermal expansion is normal (positive) from 7.4 to 300~K\@.  Furthermore, we determined from neutron diffraction measurements on a single crystal that long-range AFM order is not present at 3~K for either A-type or G-type ordering, which complements our previous determination that such ordering also does not occur at the propagation vector for stripe-type AFM ordering.\cite{Jayasekara2013}

The $\rho(T)$ and ARPES measurements and the LDA band structure calculations consistently demonstrate that SrCo$_2$As$_2$ is metallic with electron Fermi surfaces centered at $\Gamma$~(0,0) and on the edges of the first Brillouin zone (BZ) at ${\bf k}= (\pm\pi,0)$ and ${\bf k}= (0,\pm\pi)$.  An anomalous flat band with $d_{x^2-y^2}$ character is also present in the second BZ that extends over a large fraction of the zone and results in a sharp and high peak in the electronic density of states ${\cal D}(E\approx E_{\rm F})$.  This feature may be responsible for some of the anomalous properties we observe such as the negative thermal expansion and perhaps some of the $T$-dependent magnetic properties.  The $C_{\rm p}(T)$ and $\frac{1}{T_1}(T)$ NMR data also exhibit metallic Fermi liquid character via the presence of a Sommerfeld electronic contribution $\gamma T$ and a Korringa contribution, respectively.  A comparison of the measured and band structure values of $\gamma$ indicates a value of the combined many-body electron-phonon and electron-electron density of states enhancements of $\approx 45$\%.

The $\chi(T)$ data show a Curie-Weiss-like $T$ dependence above 200~K with a large negative (AFM) Weiss temperature $\theta_{\rm p}\sim -150$~K indicative of dominant AFM correlations in the system, consistent with our inelastic neutron scattering measurements that demonstrated the presence of strong stripe-type AFM fluctuations\cite{Jayasekara2013} that occur at the same wave vector as in the $A{\rm Fe_2As_2}$ high-$T_{\rm c}$ parent compounds.\cite{Johnston2010, Lumsden2010}  However, our ARPES measurements and band structure calculations for SrCo$_2$As$_2$ show no obvious Fermi surface nesting occurring at this wave vector, in contrast to the clear nesting at this wave vector in the semimetallic $A{\rm Fe_2As_2}$ compounds.\cite{Johnston2010}  The anisotropy of $1/T_1$ determined from our NMR measurements is consistent with either stripe-type AFM spin correlations or FM spin correlations or a combination of these.  At low temperatures, the $\chi(T)$ data for the two field directions show an upturn, but our NMR Knight shift measurements of the spin susceptibility show that these upturns are not intrinsic.

We modeled our $\chi(T)$ data at high $T$ by a Curie-Weiss law and obtained values for the Curie constant and for the Weiss temperature $\theta_{\rm p}\sim-150$~K, which is strongly antiferromagnetic.  We also modeled the data, including the broad maximum at 115~K, using a local-moment interconfigurational fluctuation model between a ground state level with spin $S=0$ and an excited triplet state with $S=1$, and obtained good fits to the powder-averaged $\chi(T)$ data.  We further modeled the data in terms of the $J_1$-$J_2$ frustrated local-moment spin-1/2 square lattice Heisenberg model. For the latter model, a good fit to the $\chi_{ab}(T)$ data was obtained for a FM (negative) $J_1$, AFM $J_2$ and $J_2/J_1=-1$, as suggested by the inelastic neutron scattering results,\cite{Jayasekara2013} with $|J_1|/k_{\rm B} = J_2/k_{\rm B} = 138$~K\@.

The $\chi$ for SrCo$_2$As$_2$ has a large magnitude, roughly a factor of 3--5 larger than those of the $A{\rm Fe_2As_2}$ compounds in the paramagnetic state.\cite{Johnston2010} Within an itinerant magnetism picture, this suggests an even larger Stoner enhancement than in the FeAs-based compounds.\cite{Johnston2010}  Moreover, the Wilson ratio for SrCo$_2$As$_2$ is $\approx 3.5$ and the Stoner criterion indicates that this compound should be an itinerant ferromagnet.  A similar situation occurs in BaCo$_2$As$_2$, where it was suggested that strong quantum fluctuations occur due to proximity to a quantum critical point that suppress long-range FM order.\cite{Sefat2009}

Thus, in SrCo$_2$As$_2$, both FM and stripe-type AFM fluctuations/correlations appear to coexist, as also indicated by static $\chi({\bf q})$ calculations that show peaks at both wave vectors.\cite{Jayasekara2013}  This competition, the formal intermediate valence of the Co ions and the resultant quantum fluctuations may suppress the occurrence of any type of static long-range magnetic order in this compound.  As discussed in Sec.~\ref{Sec:GndState}, we suggest that metallic SrCo$_2$As$_2$ has a gapless quantum spin-liquid (QSL) ground state.  It will be very interesting to study in detail the spin excitation spectrum of this compound and compare it with those of insulating local-moment systems that are the traditional candidates for QSL ground states.

The FeAs-based and cuprate high-$T_{\rm c}$ superconductors have a common phenomenology in which long-range AFM order of the parent compounds must be largely suppressed before superconductivity appears, but where strong AFM fluctuations must still be present to act as the superconducting glue.  Even though we have demonstrated that these two conditions are met in undoped SrCo$_2$As$_2$, this compound is not superconducting above 1.8~K\@.  Key open  questions are therefore why superconductivity does not occur in SrCo$_2$As$_2$, what the role of the FM correlations is, and whether chemical substitutions of some type or applied pressure can induce superconductivity, perhaps even at high temperatures.

\acknowledgments

We thank V. K. Anand for numerous helpful discussions and A.~Bostwick and E. Rotenberg for excellent support at the Advanced Light Source. V.O. thanks the Ames Laboratory--USDOE for providing the opportunity to be a visiting scientist at the Laboratory and also thanks the Russian Foundation for Basic Research (No.~12-02-31814) for support.  The research at Ames Laboratory was supported by the U.S. Department of Energy, Office of Basic Energy Sciences, Division of Materials Sciences and Engineering.  Ames Laboratory is operated for the U.S. Department of Energy by Iowa State University under Contract No.~DE-AC02-07CH11358.  Use of the National Synchrotron Light Source, Brookhaven National Laboratory, was supported by the U.S. Department of Energy, Office of Basic Energy Sciences, under contract No.~DE-AC02-98CH10886.  The Advanced Light Source is supported by the Director, Office of Science, Office of Basic Energy Sciences, U.S. Department of Energy under Contract No.~DE-AC02-05CH11231.

\end{document}